\documentclass[useAMS,usenatbib]{mn2e}
\usepackage[dvips]{graphicx}
\usepackage{psfig}

\title[Searching for Compton-thick AGNs at $z \sim 0.1$]
  {Searching for Compton-thick active galactic nuclei at $z \sim 0.1$}
\author[A.D.~Goulding et al.]
  {A.D.~Goulding$^1$; D.M.~Alexander$^1$; J.R.~Mullaney$^1$; J.M.~Gelbord$^{1,2}$; R.C.~Hickox$^1$;
\newauthor
   M.~Ward$^1$ \& M.G.~Watson$^3$ \\
  $^1$Department of Physics, Durham University, South Road, Durham. \\
  $^2$Department of Astronomy and Astrophysics, 525 Davey Laboratory, The Pennsylvania State University, University Park, PA 16802. \\
  $^3$Department of Physics \& Astronomy, University of Leicester, Leicester, LE1 7RH, UK}
\date{Released 2010 Xxxxx XX}

\pagerange{\pageref{firstpage}--\pageref{lastpage}} \pubyear{2010}

\def\LaTeX{L\kern-.36em\raise.3ex\hbox{a}\kern-.15em
    T\kern-.1667em\lower.7ex\hbox{E}\kern-.125emX}

\def\cm{{\rm\thinspace cm}}

\def\erg{{\rm\thinspace erg}}

\def\km{{\rm\thinspace km}}

\def\Lsun{\hbox{$\rm\thinspace L_{\odot}$}}

\def\Mpc{{\rm\thinspace Mpc}}
\def\Msun{\hbox{$\rm\thinspace M_{\odot}$}}

\def\s{{\rm\thinspace s}}

\def\ergps{\hbox{$\erg\s^{-1}\,$}}

\def\kmps{\hbox{$\km\s^{-1}\,$}}
\def\kmpspMpc{\hbox{$\km\s^{-1}\Mpc^{-1}\,$}}

\def\pcmsq{\hbox{$\cm^{-2}\,$}}
\def\Mbh{\hbox{${\rm M}_{\rm BH}$}}

\def\um{\hbox{$\ \umu {\rm m}$}}

\def\nev{{[Ne{\sc v}]~}}
\def\oiv{{[O{\sc iv}]~}}
\def\oiii{{[O{\sc iii}]~}}

\begin{document}

\label{firstpage}

\maketitle

\begin{abstract}
Using a suite of X-ray, mid-infrared (IR) and optical active galactic
nuclei (AGN) luminosity indicators, we search for Compton-thick AGNs
with intrinsic $L_{\rm X} > 10^{42} \ergps$ at $z \sim 0.03$--0.2, a
region of parameter space which is currently poorly constrained by
deep narrow-field and high-energy ($E>10$~keV) all-sky X-ray
surveys. We have used the widest {\it XMM-Newton} survey (the
serendipitous source catalogue) to select a representative sub-sample
(14; $\approx 10$ percent) of the 147 X-ray undetected candidate
Compton-thick AGNs in the Sloan Digital Sky Survey (SDSS) with $f_X /
f_{\rm [OIII]} < 1$; the 147 sources account for $\approx 50$~percent
of the overall Type-2 AGN population in the SDSS--XMM overlap
region. We use mid-IR spectral decomposition analyses and
emission-line diagnostics, determined from pointed {\it Spitzer}-IRS
spectroscopic observations of these candidate Compton-thick AGNs, to
estimate the intrinsic AGN emission (predicted 2--10~keV X-ray
luminosities, $L_X \approx (0.2$--$30) \times 10^{42} \ergps$). On the
basis of the optical [O{\sc iii}], mid-IR \oiv and $6 \um$ AGN
continuum luminosities we conservatively find that the X-ray emission
in at least 6/14 ($\goa 43$ percent) of our sample appear to be
obscured by Compton-thick material with $N_H > 1.5 \times 10^{24}
\pcmsq$. Under the reasonable assumption that our 14 AGNs are
representative of the overall X-ray undetected AGN population in the
SDSS--XMM parent sample, we find that $\goa 20$~percent of the optical
Type-2 AGN population are likely to be obscured by Compton-thick
material. This implies a space-density of log$(\Phi) \goa
-4.9$~Mpc$^{-3}$ for Compton-thick AGNs with $L_X \goa 10^{42} \ergps$
at $z \sim 0.1$, which we suggest may be consistent with that
predicted by X-ray background synthesis models.  Furthermore, using
the $6 \um$ continuum luminosity to infer the intrinsic AGN luminosity
and the stellar velocity dispersion to estimate $\Mbh$, we find that
the most conservatively identified Compton-thick AGNs in this sample
may harbour some of the most rapidly growing black holes (median $\Mbh
\approx 3 \times 10^7 \Msun$) in the nearby Universe, with a median
Eddington ratio of $\eta \approx 0.2$.
 \end{abstract}
 


\section{Introduction}
There is now strong observational evidence that all massive galaxies
($M_* \approx 10^{10}$--$10^{12} \Msun$) in the nearby Universe host a
central supermassive black hole (SMBH; $\Mbh \approx 10^6$--$10^9
\Msun$; \citealt{kormendy95}). These SMBHs have grown through mass
accretion events (e.g., \citealt{soltan82}; \citealt{rees84}), during
so-called active galactic nucleus (AGN) phases. The seminal discovery
that the masses of SMBHs are proportional to those of their stellar
spheroids implies a strong physical association between AGN activity
and galaxy evolution (e.g., \citealt{magorrian98};
\citealt{ferrarese00}; \citealt{gebhardt00}; \citealt{tremaine02}). To
fully interpret the role played by AGN in this symbiosis requires a
complete census of obscured and unobscured AGNs across cosmic time.

Unbiased deep and wide-field X-ray surveys have been instrumental in
the identification of a large proportion of the AGN population to high
redshifts ($z \sim 5$; e.g.,
\citealt{dma01,barger03,fiore03,tozzi06,brusa10}). Using the
exceptional sensitivities of {\it XMM-Newton} and the {\it Chandra
  X-ray Observatory}, $> 80$ per cent of the X-ray background (XRB)
has been resolved into discrete sources at soft energies (0.5--5~keV;
e.g., \citealt{worsley05, hickox06, hickox07a}). However, AGN
synthesis models for the XRB predict that $\sim 50$ per cent of the
AGN population may be heavily obscured and remains undetected at $E >
6$~keV in deep X-ray surveys (e.g., \citealt{gilli07, treister09}).

Using high-quality X-ray spectroscopic analyses of objects in the
local Universe, it is now well-established that the majority of AGNs
are obscured along the line-of-sight by large columns of gas and dust
(e.g., \citealt{risaliti99, matt00}). The presence of this obscuring
material results in strongly depressed nuclear X-ray emission observed
at $E \sim 0.5$--10~keV. For those AGNs with column densities
exceeding the inverse Thomson cross-section ($N_H \sim 1.5 \times
10^{24} \pcmsq$; i.e., Compton-thick absorption) very few photons are
detected at $E<10$~keV due to significant absorption and
scattering. Moreover, for sources with $N_H > 10^{25} \pcmsq$, the
entire high energy spectrum is down-scattered and, eventually,
absorbed by the heavily Compton-thick material. Consequently, the
observed X-ray flux in Compton-thick AGNs is often rendered so weak
that it becomes comparable to the X-ray emission arising from the
host-galaxy, making their detection extremely difficult. The direct
identification of mildly Compton-thick AGNs ($N_H \sim (1.5$--$10)
\times 10^{24} \pcmsq$) is possible through X-ray observations at $E >
10$~keV (e.g., using {\it Beppo}-SAX, {\it Swift}, {\it Suzaku})
where the relatively unabsorbed high-energy emission can be
detected. However, the sensitivities of current $E>10$~keV
observatories are substantially limited by high backgrounds, poor
effective areas and inadequate spatial resolutions. Indeed, to date,
only 18 Compton-thick AGNs have been unambiguously identified in the
Universe at $E > 10$~keV, mainly at $z \loa 0.01$ (for a recent review,
see \citealt{dellaceca08}).

In the absence of higher-energy $E>10$~keV data, the presence of a
Compton-thick AGN may still be inferred using indirect methods: (1)
from the detection of a high equivalent width ($> 1$~keV) Fe
K$_\alpha$ fluorescence line at $E \sim 6.4$~keV (e.g.,
\citealt{awaki91}), and/or (2) nuclear emission which has been
reflected into the line-of-sight by the highly ionised optically-thick
material (a so-called, Compton-reflection component; e.g.,
\citealt{ghisellini94,matt96}). However, the detection of either of
these Compton-thick AGN signatures is still difficult due to the
required high sensitivity of the X-ray data (spectra containing $\goa
200$~counts). For example, given the faint X-ray fluxes, even at low
redshifts ($z \sim 0.05$), long exposure times with the most sensitive
X-ray observatories (of the order 100s of kiloseconds with {\it
  Chandra} and {\it XMM-Newton}) are required to detect FeK$_\alpha$
at a high significance. Indeed, only a further $\approx 30$ local ($z
< 0.01$) AGNs have been robustly determined to be Compton-thick AGNs
in the absence of $E > 10$~keV data (see Comastri 2004;
\citealt{dellaceca08} and references there-in). Hence, although
Compton-thick AGNs are predicted to comprise a large proportion of the
overall AGN population ($\goa 40$~percent;
\citealt{risaliti99,matt00}), to date, only $\approx 50$ Compton-thick
AGNs have been robustly identified in the nearby Universe at $z \loa
0.05$ \citep{comastri04, dellaceca08}; conversely, only $\approx 30$
candidate Compton-thick AGNs (i.e., those with high EW FeK or
reflection-dominated spectra) have been identified in high redshift
X-ray surveys (e.g., \citealt{Norman02,tozzi06,georgantopoulos09}).

Given the required X-ray sensitivity to directly identify
Compton-thick AGNs using X-ray data alone, only a small fraction of
the population can be discovered using current instrumentation. In
recent years, new techniques have been developed to discover
Compton-thick AGN candidates using complimentary wide-field optical
surveys with pointed mid-infrared (IR) observations, allowing us to
probe $\approx 2$--3 orders of magnitude lower in the $z$--$L_X$ plane
than using X-ray data alone. These approaches are promising since the
reprocessed mid-IR continuum emission and high-excitation optical
and mid-IR narrow-line emission (i.e., \oiii $\lambda 5007$; \nev
$14.32\um$; \oiv $25.89 \um$) in AGN are relatively unaffected by the
optically-thick X-ray obscuring material in the central region and,
therefore, provide reliable measurements of the intrinsic luminosity
of even the most heavily Compton-thick AGNs (e.g., \citealt{heckman05,
  panessa06, melendez08b, diamond09, goulding10}). For example,
through examination of a local optically-selected AGN sample,
\citet{maiolino98} and \citet{bassani99} find that those AGNs with
X-ray--\oiii flux ratios of $f_X/f_{\rm [OIII]} < 1$ almost invariably
host intrinsically obscured central sources (many of which are Compton
thick), and those with $f_X/f_{\rm [OIII]} < 0.1$ always appear to be
Compton thick (see also \citealt{akylas09}; hereafter,
AG09). Furthermore, this diagnostic has successfully identified new
Compton-thick AGNs which have since been unambiguously confirmed using
high-quality X-ray data (e.g., NGC~5135; \citealt{levenson04}).

Clearly, indirect AGN luminosity indicators provide good first-order
approximations as to whether an AGN is Compton thick. Greater
reliability in identifying Compton-thick AGNs can therefore be made
when considering multiple diagnostics, particularly those which probe
different regions of the AGN (e.g., the emission line region and the
reprocessed continuum emission). Recently, \citet{vignali10} combined
pointed {\it Chandra}-ACIS observations with optical emission-line
{\it and} mid-IR continuum luminosities to identify six Compton-thick
quasars ($L_{\rm [OIII]} > 2 \times 10^9 \Lsun$) at $z \sim
0.40$--0.73 in the Sloan Digital Sky Survey (SDSS). Somewhat similar
approaches have also been adopted by
\citet{dma08,lamassa09,bauer10,donley10} using {\it Spitzer} IR
spectroscopy and/or optical spectroscopy to identify high-redshift
X-ray undetected Compton-thick AGNs in deep and wide-field
surveys. Whilst each of these studies have successfully identified
Compton-thick AGNs using multi-wavelength analyses, they sample only
the most luminous systems ($L_{\rm X,intr} \goa 10^{44} \ergps$) where
the predicted space-density of Compton-thick AGNs, even at $z \sim 2$,
is relatively low ($\phi \loa 10^{-5}$~Mpc$^{-3}$;
\citealt{gilli07}). In order to clearly understand the evolution of
these Compton-thick sources, it is vital to also identify the more
modest luminosity population ($L_{\rm X,intr} \approx [0.1$--$1]
\times 10^{43} \ergps$), which comprise the most energetically
dominant AGNs in the nearby Universe ($z \sim 0.1$; e.g.,
\citealt{ueda03,Ebrero09,Aird10}).

In this paper, we identify a sample of nearby ($z \sim 0.03$--0.2)
X-ray undetected optically selected candidate Compton-thick AGNs from
a large cosmological volume which is well-matched to that of the {\it
  Chandra} Deep Fields (CDFs; \citealt{giacconi02, dma03a,Luo08}) at
$z \sim 0.5$--2.5 (comoving volume of $V \approx 4.6 \times
10^{6}$~Mpc$^{3}$). We use pointed high signal-to-noise {\it Spitzer}
IR spectroscopy and $24\um$ photometry combined with the unprecedented
wide-field coverage of the SDSS to explore the ubiquity of typical
Compton-thick AGNs at $z \sim 0.1$. In section 2, we outline the
construction of our sample of 14 optical narrow-line AGNs derived from
the population of galaxies in the SDSS which lie in the large overlap
region with the {\it XMM-Newton} Serendipitous Survey ($\approx
100$~deg$^{2}$). These AGNs are all undetected to faint flux limits in
$E \sim 2$--12~keV {\it XMM-Newton} observations, and based on their
X-ray--\oiii flux ratio limits are likely to be heavily obscured (and
possibly Compton thick). In section 3, we discuss the data reduction
techniques for our {\it Spitzer} observations and present mid-IR
AGN--starburst spectral decompositions to understand both the
properties of the host galaxies and the central source in these
obscured AGNs. In section 4, we use mid-IR narrow-line emission and
AGN-produced mid-IR continuum emission to determine the intrinsic
luminosity of the obscured AGNs. We combine each of these AGN
luminosity indicators in order to reliably identify which sources are
Compton-thick AGNs. We use these results to further constrain the
ubiquity of Compton-thick AGNs at $z \sim 0.1$. Throughout, we adopt a
standard $\Lambda$CDM cosmology of $H_0 = 71 \kmpspMpc$, $\Omega_M =
0.30$, and $\Omega_\Lambda = 0.70$.


\begin{figure}
\includegraphics[width=1.00\linewidth]{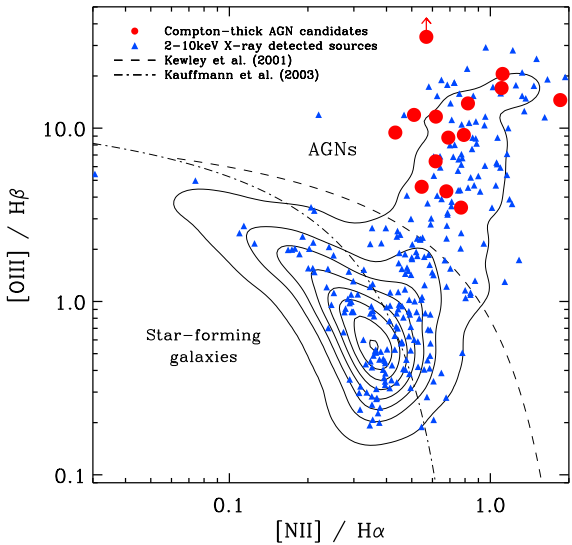}
\caption{Optical emission line diagnostic diagram presenting the 2690
  galaxies detected in the seventh data release of the Sloan Digital
  Sky Survey with serendipitous X-ray coverage from {\it XMM-Newton}
  at $0.03 < z < 0.2$ (solid contours). The fourteen X-ray undetected
  optical AGNs selected for our sample are shown with red solid
  circles. The 272 galaxies with hard-band X-ray detections ($E >
  2$~keV) are shown with blue solid triangles. The empirical H{\sc ii}
  star-forming classification curve presented by Kauffmann et
  al. (2003) and the extreme starburst line of \citet{kewley_bpt} are
  also shown with dash-dotted and dashed curves, respectively.}
\label{fig:bpt}
\end{figure}

\section{Sample Selection}
\label{sec:g10b_sampleselect}

We select our candidate Compton-thick AGN sample on the basis of their
optical and X-ray properties. Sources that are identified to be AGNs
using traditional optical emission line diagnostics (e.g.,
\citealt{bpt}) but are undetected to faint limits in wide-field {\it
  XMM-Newton} observations (i.e., $f_X/f_{\rm [OIII]} < 1$) are strong
candidates for containing heavily obscured AGNs (e.g.,
\citealt{bassani99, panessa02, akylas09}).\footnote{Throughout this
  manuscript we define {\it X-ray undetected} as those sources
  undetected in the hard band ($E>2$~keV) unless otherwise stated.}
Here we provide the details behind the construction of our sample of
X-ray undetected optically identified AGNs (i.e., candidate
Compton-thick AGNs).

\subsection{Construction of the optical--X-ray catalogue}

We construct a parent sample of all optical spectroscopically
identified galaxies in the $\approx 100$~deg$^2$ overlap region
between the seventh data release of the SDSS (\citealt{sdss_dr7};
hereafter SDSS-DR7) and the second source catalogue of the {\it
  XMM-Newton} Serendipitous survey (\citealt{watson09}; hereafter
2XMMi). We define the redshift range for our shallow wide-area sample
based on the combined available cosmological volume in the deep 2Ms
``pencil-beam'' CDF-North ($\approx 448$~arcmin$^2$; \citealt{dma03a})
and CDF-South ($\approx 436$~arcmin$^2$; \citealt{giacconi02,Luo08})
surveys. At $z \sim 0.5$--2.5, where the CDFs are complete towards
X-ray luminous AGNs, the encompassed comoving volume is $V \sim 4.57
\times 10^6$~Mpc$^3$, which is equivalent to the comoving volume in
the redshift range of $z \sim 0.03$--0.2 in our SDSS-2XMMi selected
sample.

\subsubsection{AGN identification in the SDSS-DR7}
The SDSS-DR7 is currently the largest publicly available optical
spectroscopic catalogue ($\approx 9830$~deg$^{2}$) containing 929,555
spectroscopic source redshifts.\footnote{The SDSS-DR7 data archive
  server is available at http://www.sdss.org/dr7/} Previous studies
have used past data releases of the survey to show that through
careful spectral analyses, general galaxy and AGN properties can be
derived from these large datasets (e.g., \citealt{kauff03b, heckman04,
  gre_ho07}). We select all galaxies with well detected narrow
\oiii$\lambda 5007$, H$\alpha$, [N{\sc ii}]$\lambda 6585$
emission-lines (S/N~$> 5$).\footnote{SDSS spectra are obtained through
  3 arcsecond fibers; at the median redshift of our sample ($z \sim
  0.08$) this projected aperture is equivalent to a physical region of
  $\approx 5$~kpc, and hence encompasses all of the narrow-line region
  emission as well as a large fraction of the host galaxy emission.}
All galaxies with detected broad Balmer emission lines (here defined
as a full-width half maximum $> 700\kmps$) are removed as these
sources are unlikely to be intrinsically obscured by a gas/dust-rich
geometrically thick torus. AGNs which are heavily obscured are often
found to be hosted in dust-rich galaxies, and thus are likely to be
strongly reddened (i.e., H$\alpha$--H$\beta$ ratios $\gg 3.1$; e.g.,
\citealt{GA09}). Hence, whilst useful in unambiguously discriminating
between the properties of galaxies (e.g., Kauffmann et~al. 2003;
\citealt{wild10}), we purposely do not limit our selection to only
galaxies with well-detected H$\beta$ emission. Sources are separated
by classification based on their optical emission-line ratios in a
traditional diagnostic diagram (hereafter, BPT diagram; e.g.,
\citealt{bpt}). We conservatively identify the narrow-line AGNs in the
SDSS-DR7 as those which lie above the theoretical starburst limit of
\cite{kewley_bpt}. See Fig.~\ref{fig:bpt}.

\subsubsection{SDSS AGNs in the {\it XMM-Newton} footprint}

The 2XMMi catalogue identifies all X-ray sources detected in the 3491
observations made during the first $\approx 8$ years of {\it
  XMM-Newton} operations (Watson et~al. 2009). Its unprecedented sky
coverage (360 deg$^{2}$) and sensitivity (median {\it XMM-Newton}
exposures of 20--50~ks) currently provides an exceptional resource for
the unbiased identification of obscured AGN activity throughout the
Universe. Using an automated reduction and analysis pipeline, the
2XMMi catalogue provides source positions, exposure times, X-ray
fluxes and band ratios of all detected sources which serendipitously
fall within the field-of-view of previous {\it XMM-Newton}
observations.

All sources in our SDSS parent sample are matched to 2XMMi using a 3.7
arcsecond radius, which is chosen as a good compromise between
maximising source numbers and minimising the probability of spurious
matches (e.g., Watson et~al. 2009). The matching algorithm is
restricted to sources within 14 arc-minutes of the aim point of each
{\it XMM-Newton} observation to minimise the likelihood of spurious
matches due to the degradation of the X-ray PSF far off-axis. Based on
the X-ray/optical positional analysis of SDSS quasars and the 2XMMi
catalogue by \citet{watson09}, if we assume no systematic offsets,
then we expect our XMM--SDSS matching to be $\approx 92$~percent
complete. We identify all optical sources in the 2XMMi which have $3
\sigma$ detections at $E \sim 2$--12~keV using the PN detector. For
all other matched sources (i.e., those which lie within the footprint
of an {\it XMM-Newton} observation but are undetected in the hard-band
of the 2XMMi catalogue), we use {\sc flix} to compute robust $3\sigma$
(likelihood threshold of 6.6) X-ray upper-limits in this
band.\footnote{{\sc flix} is a purpose-built program provided by the
  {\it XMM-Newton} Survey Science Center. It provides robust estimates
  of the X-ray upper limit to a given point in the sky for a source
  which has not been detected in the 2XMMi catalogue. For a discussion
  of the upper limit algorithm see \citet{carrera07} and for further
  documentation see http://www.ledas.ac.uk/flix/flix\_help.html.}
Using a sub-sample of the matched sources that are formally undetected
in the hard band in 2XMMi, we tested the use of {\sc flix} to provide
X-ray upper-limits. Broadly, we find that the upper limits provided by
{\sc flix} are consistent with the fluxes within $\pm 3 \sigma$ given
by 2XMMi. For the sake of comparison with previous studies, we convert
these 2--12~keV upper-limits to 2--10~keV limits assuming a powerlaw
spectrum with spectral index of $\Gamma = 1.4$ where $F_{\nu} \propto
\nu^{-(\Gamma -1)}$; $\Gamma =1.4$ is the spectral slope of the X-ray
background and similar to that of many absorbed AGNs. Our final
combined parent sample of SDSS-DR7 galaxies at $z \sim 0.03$--0.2 with
complimentary hard X-ray {\it XMM-Newton} coverage is 2690 objects
(272 are hard X-ray detected sources). The median redshift of the
sample is $\approx 0.09$. Of these galaxies, 334 ($\approx
12$~percent) lie above the theoretical starburst limit and are
classified as optical narrow-line (NL) AGNs (i.e., 101 are X-ray
detected and 233 are X-ray undetected AGNs; see Fig.~\ref{fig:bpt}).

\subsubsection{A sample of candidate Compton-thick AGNs}
\label{sec:sample-cand-compt}

Assuming the optical emission-lines and X-ray AGN emission are
well-correlated (e.g., \citealt{mulchaey94, alonso97}), sources which
are optically classified as AGNs but are undetected to faint limits in
relatively deep X-ray observations are likely to be those with heavily
attenuated X-ray emission, similar to the objects currently missed in
deep X-ray surveys. From our well-defined parent sample of 334 optical
NL AGNs, 233 are not detected in the hard-band of the 2XMMi
catalogue. In this section, we outline the selection method for our
sample of 14 hard-band undetected candidate Compton-thick AGNs.

\begin{figure}
\includegraphics[width=1.00\linewidth]{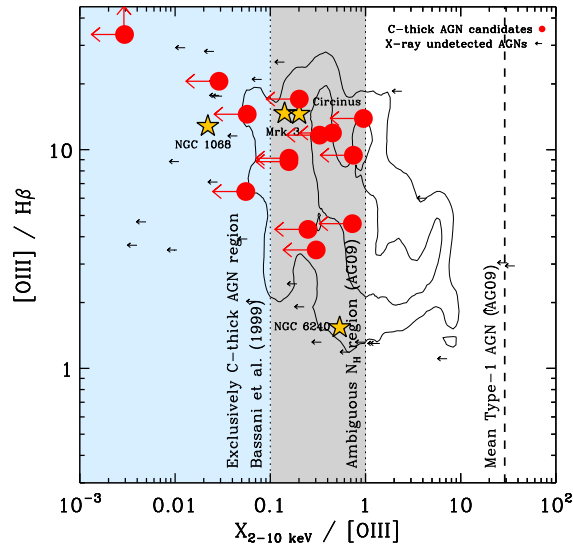}
\caption{Intrinsic obscuration optical--X-ray diagnostic diagram for
  AGNs. Contours indicate the distribution of all optical AGNs in the
  SDSS-DR7 which lie in the {\it XMM-Newton} footprint (i.e., those
  galaxies which lie above the Kewley et al. 2001 extreme starburst
  line presented in Fig.~\ref{fig:bpt}). These contours enclose 50\%
  and 90\% of the whole X-ray undetected parent sample; outliers are
  shown with small arrows. We select 14 candidate Compton-thick AGNs
  for {\it Spitzer} observations based on their $f_X / f_{\rm [OIII]}$
  ratio (filled circles). Four of these sources lie in the region
  exclusively occupied by Compton-thick AGNs ($f_X / f_{\rm [OIII]} <
  0.1$; e.g., Bassani et~al. 1999), and 10 are selected from the
  heavily obscured $N_H$ region ($f_X / f_{\rm [OIII]} \sim 0.1$--1.0)
  which is often found to contain Compton-thick AGNs (e.g., Bassani
  et~al. 1999; AG09). We also show the mean Type-1 (i.e., unobscured)
  AGN $f_X / f_{\rm [OIII]}$ ratio from AG09 and for comparison, four
  well-studied local `bona-fide' Compton-thick AGNs (Circinus, Mrk~3,
  NGC~1068 and 6240; data is taken from Bassani et~al. 1999; stars).}
\label{fig:XO3}
\end{figure}

\begin{table*}
\begin{minipage}{175mm}
\begin{center}
\setlength{\tabcolsep}{1.0mm}
\caption{Basic source properties of the candidate Compton-thick AGNs}
\label{tab:srce_props}
\begin{tabular}{lllrccccccccc}
\hline
  \multicolumn{1}{|c|}{ID} &
  \multicolumn{1}{|c|}{Source name} &
  \multicolumn{1}{|c|}{$\alpha_{J2000}$} &
  \multicolumn{1}{|c|}{$\delta_{J2000}$} &
  \multicolumn{1}{|c|}{$z$} &
  \multicolumn{1}{|c|}{$D_{\rm L}$} &
  \multicolumn{1}{|c|}{log($M_{\rm BH}$)} &
  \multicolumn{1}{|c|}{\underline{\oiii}} &
  \multicolumn{1}{|c|}{\underline{[NII]}} &
  \multicolumn{1}{|c|}{\underline{H$\alpha$}} & 
  \multicolumn{1}{|c|}{$A_V$} &
  \multicolumn{1}{|c|}{log($L_{\rm [OIII]}$)} &
  \multicolumn{1}{|c|}{log($L_{\rm HX}$)} \\
  \multicolumn{1}{|c|}{} &
  \multicolumn{1}{|c|}{} &
  \multicolumn{1}{|c|}{(deg)} &
  \multicolumn{1}{|c|}{(deg)} &
  \multicolumn{1}{|c|}{} &
  \multicolumn{1}{|c|}{(Mpc)} &
  \multicolumn{1}{|c|}{($\Msun$)} &
  \multicolumn{1}{|c|}{H$\beta$} &
  \multicolumn{1}{|c|}{H$\alpha$} &
  \multicolumn{1}{|c|}{H$\beta$} &
  \multicolumn{1}{|c|}{(mags)} &
  \multicolumn{1}{|c|}{($\ergps$)} &
  \multicolumn{1}{|c|}{($\ergps$)} \\
  \multicolumn{1}{|c|}{(1)} &
  \multicolumn{1}{|c|}{(2)} &
  \multicolumn{1}{|c|}{(3)} &
  \multicolumn{1}{|c|}{(4)} &
  \multicolumn{1}{|c|}{(5)} &
  \multicolumn{1}{|c|}{(6)} &
  \multicolumn{1}{|c|}{(7)} &
  \multicolumn{1}{|c|}{(8)} &
  \multicolumn{1}{|c|}{(9)} &
  \multicolumn{1}{|c|}{(10)} &
  \multicolumn{1}{|c|}{(11)} &
  \multicolumn{1}{|c|}{(12)} &
  \multicolumn{1}{|c|}{(13)} \\
\hline
1  & SDSS~J094046$+$033930 & 145.19287 & $3.65839$  & 0.08730 & 392.8 & $7.94$  & 20.55    & 1.11 & 16.61    & 4.86    & 42.94 & $<41.40$ \\
2  & SDSS~J094506$+$035551 & 146.27664 & $3.93087$  & 0.15593 & 733.6 & $7.68$  & 11.93    & 0.51 & 4.68     & 1.20    & 42.73 & $<42.39$ \\
3  & SDSS~J100328$+$554154 & 150.86636 & $55.69831$ & 0.14602 & 682.8 & $7.53$  & 13.89    & 0.82 & 4.72     & 1.22    & 42.39 & $<42.36$ \\
4  & SDSS~J101757$+$390528 & 154.48708 & $39.09110$ & 0.05392 & 237.0 & $7.29$  & 4.59     & 0.55 & 4.94     & 1.35    & 41.33 & $<41.19$ \\
5  & SDSS~J102142$+$130550 & 155.42455 & $13.09733$ & 0.07650 & 341.7 & $6.93$  & 4.32     & 0.68 & 9.69     & 3.30    & 41.83 & $<41.23$ \\
6  & SDSS~J111521$+$424217 & 168.83784 & $42.70484$ & 0.19714 & 950.9 & -       & $>33.63$ & 0.57 & $>33.30$ & $>6.88$ & 44.68 & $<42.15$ \\
7  & SDSS~J115658$+$550822 & 179.24117 & $55.13932$ & 0.07980 & 357.2 & $7.72$  & 17.04    & 1.10 & 7.40     & 2.52    & 42.96 & $<41.62$ \\
8  & SDSS~J121355$+$024753 & 183.47861 & $2.79806$  & 0.07429 & 331.3 & $6.77$  & 6.44     & 0.62 & 6.30     & 2.05    & 42.49 & $<41.23$ \\
9  & SDSS~J123026$+$414258 & 187.60876 & $41.71606$ & 0.12486 & 576.0 & $7.50$  & 8.82     & 0.70 & 8.53     & 2.93    & 42.58 & $<41.77$ \\
10 & SDSS~J134133$-$002432 & 205.38905 & $-0.40892$ & 0.07173 & 319.3 & $<8.30$ & 11.66    & 0.62 & 6.34     & 2.07    & 42.07 & $<41.60$ \\
11 & SDSS~J135858$+$651546 & 209.61739 & $65.26288$ & 0.03259 & 141.1 & $6.85$  & 3.48     & 0.77 & 9.78     & 3.32    & 41.48 & $<40.96$ \\
12 & SDSS~J142931$+$425149 & 217.37810 & $42.86364$ & 0.15500 & 728.8 & $7.60$  & 9.14     & 0.79 & 12.09    & 3.94    & 42.32 & $<42.16$ \\
13 & SDSS~J215650$-$074533 & 329.20630 & $-7.75905$ & 0.05541 & 243.8 & $7.14$  & 9.43     & 0.43 & 4.41     & 1.02    & 41.61 & $<41.49$ \\
14 & SDSS~J221742$+$000908 & 334.42340 & $0.15209$  & 0.04514 & 197.2 & $6.36$  & 14.53    & 1.85 & 11.15    & 3.71    & 41.80 & $<40.56$ \\

  \hline\end{tabular}
\medskip
\end{center}

{\tiny NOTES:} (1) Source-identification; (2) SDSS source name; (3--4)
J2000 positional co-ordinates from the SDSS-DR7; (5) SDSS
spectroscopic redshift; (6) luminosity distance in megaparsecs
calculated using our adopted cosmology; (7) logarithm of black hole
mass in units of solar masses derived from the stellar velocity
dispersion (MPA-JHU DR7 release) using the $M$--$\sigma$ relation
\citep{gebhardt00}; (8--10) Emission-line ratios from SDSS-DR7; (11)
Implied $A_V$ in magnitudes derived from the Balmer decrement
($H\alpha / H\beta$; see section~\ref{sec:sample-cand-compt}); (12)
Logarithm of dust extinction corrected [O{\sc iii}] luminosity; (13)
Logarithm of the 3-$\sigma$ upper limit of the 2-10 keV X-ray
luminosity derived from fluxes produced using {\sc flix} (see footnote
4).
\end{minipage}
\end{table*}

The de-reddened \oiii luminosity is assumed to be a good tracer of AGN
power (e.g., \citealt{heckman05, netzer06, panessa06}). In order to
identify obscured AGN candidates, we follow \citet{maiolino98} and
\citet{bassani99} by using the flux ratio of de-reddened (intrinsic)
\oiii and observed (absorbed) 2--10 keV X-ray emission, and compare it
to the \oiii/H$_{\beta}$ ratio in a new diagnostic diagram analogous
to a BPT diagram; see Fig.~\ref{fig:XO3}. Optical luminosities are
corrected for dust-reddening towards the AGN NL-region using the
Balmer decrement (i.e., the observed H$\alpha$--H$\beta$ ratio;
\citealt{ward87}), an intrinsic ratio of 3.1 \citep{oster_book06} and
a standard $R=3.1$ \citet{cardelli89} extinction curve. The parent AGN
sample are found to have de-reddened \oiii luminosities in the range
$L_{\rm [OIII]} \approx (0.03$--$500) \times 10^{42} \ergps$ (median
$L_{\rm [OIII]} \approx 10^{42} \ergps$). AG09 find that the average
X-ray--\oiii flux ratio for unobscured AGNs (i.e., Type~1s; $N_H <
10^{22} \pcmsq$) is $\approx 30$, whilst heavily obscured AGNs
(Type~2s) typically exhibit lower values of $f_X / f_{\rm
  [OIII]}$. AGNs with $f_X / f_{\rm [OIII]} < 0.1$ are invariably
found to be Compton thick and a significant proportion of
Compton-thick AGNs have $f_X / f_{\rm [OIII]} \sim 0.1$--$1.0$ in
addition to heavily absorbed Compton-thin AGNs. One-hundred and
forty-seven ($\approx 63$~percent) of the 233 X-ray undetected AGNs
have $f_X / f_{\rm [OIII]} < 1.0$ (24 have $f_X / f_{\rm [OIII]} <
0.1$) from which we select a representative sub-sample of 14 ($\approx
10$~percent) to be further investigated using pointed mid-IR
spectroscopic and photometric observations. Our sample of 14 AGNs are
well-matched to the parent sample of X-ray undetected AGNs with a
redshift distribution of 0.03--0.2 (median $\sim 0.08$) and $L_{\rm
  [OIII]} \approx (0.2$--$500) \times 10^{42} \ergps$ (median $L_{\rm
  [OIII]} \approx 2 \times 10^{42} \ergps$). For completeness, we also
note that 11/14 of our sources are detected in at least one of the
softer X-ray bands ($E < 2$~keV) in 2XMMi. These detections, in many
cases, may be due to a scattered or reprocessed soft X-ray component
(e.g., \citealt{matt00}). Clearly, this softer component may partially
contribute to the flux at $E > 2$~keV. However, any correction which
could be made to our $3 \sigma$ upper-limits would only serve to
reduce the current hard-band limits. The basic source properties for
our sample of 14 candidate Compton-thick AGNs are shown in
Table~\ref{tab:srce_props}.


\begin{figure*}
\includegraphics[width=1.0\textwidth]{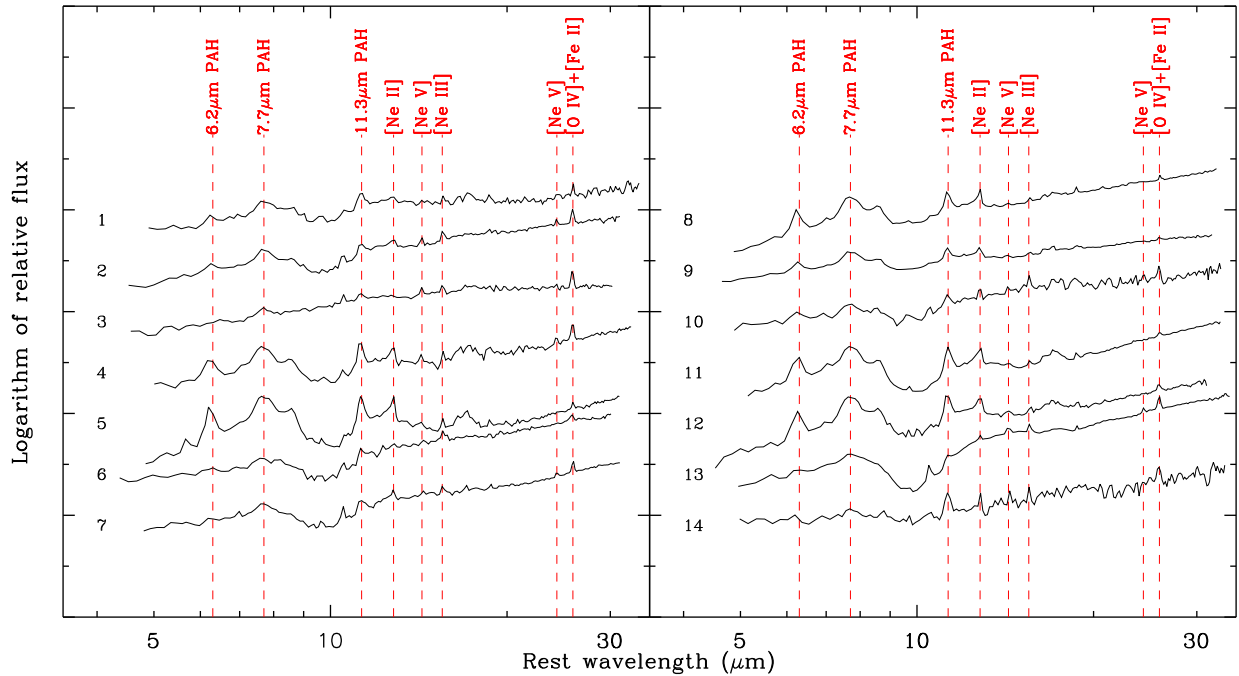}
\caption{Rest-frame low-resolution {\it Spitzer}-IRS spectroscopy of
  the 14 X-ray undetected SDSS AGNs in our sample. The prominent
  emission-line and PAH features that may be detected are
  highlighted.}
\label{fig:spec}
\end{figure*}

\section{Mid-IR Spectroscopy and Photometry}
\label{sec:midir_spec_photom}

We have used the {\it Spitzer} Infra-Red Spectrograph (IRS) and
Multi-band Imaging Photometry for {\it Spitzer} (MIPS) to observe the
14 candidate Compton-thick AGNs selected in Section 2 (PID:50818; PI:
D.Alexander). In this section, we present the reduction methodology
and resulting spectroscopy and photometry for these 14 targets. In
order to robustly assess the intrinsic luminosity of the central
sources in these AGNs, we also present mid-IR spectral decomposition
analyses to isolate the AGN continuum and star-formation emission.

\subsection{{\it Spitzer}-IRS Spectral Reduction and Analysis}
\label{subsec:spitz_irs_reduc}

Each of the 14 candidate Compton-thick AGNs were observed in spectral
staring mode with the low-resolution modules (short-low [SL;
  5.2--$14.5\um$] and long-low [LL; 14.0--$38.0\um$]; $R \approx
57$--127) of the {\it Spitzer}-IRS instrument \citep{spitzer_irs}. The
sources were observed between 30th November 2008 and 24th February
2009 using ramp durations of 60 seconds $\times$ 10 (4) cycles and 120
seconds $\times$ 4 cycles for the SL1 (SL2) and LL modules,
respectively. The total integration time for each of the sources was
0.5 hours.

The two-dimensional Basic Calibrated Data (BCDs) images produced by
the S18.7.0 {\it Spitzer} Science Center (SSC) pipeline were retrieved
and further analyzed using our custom {\sc idl} reduction routine (see
\citealt{goulding_phd10}). Individual BCDs were rigorously cleaned of
rogue `hot' pixels using a customised version of {\sc irsclean}. Next,
individual rows were fit as a function of time to remove latent charge
which exists on the detector after bright and/or long
observations. The cleaned BCDs were averaged in the differing nod
positions, which were then used to perform alternate background
subtractions of the source in each nod position.

Spectral extraction was performed using the {\it Spitzer}-IRS Custom
Extraction ({\sc spice}) software provided by the SSC. The
spectroscopy was extracted using an optimally calibrated 2-pixel wide
spectral window to maximize the signal-to-noise ratio of each
spectrum. Flux uncertainties were estimated for each of the spectra
using a second spectral window offset from the source in the spatial
direction. The spectra for each of the modules for an individual
object were corrected for their differing apertures and normalized to
the flux level of the 1st LL module. Orders were clipped of spectral
noise (see the {\it Spitzer}-IRS handbook for further information) and
stitched together by fitting low-order polynomials to produce the
final spectra.

\subsection{{\it Spitzer}-MIPS Reduction and Analysis}
\label{subsec:spitz_mips}

In order to measure accurate emission-line and continuum fluxes we use
{\it Spitzer}-MIPS $24\um$ photometry to flux calibrate the IRS
spectroscopy. Eleven of our 14 {\it Spitzer}-IRS targets were observed
with the MIPS photometer. The remaining three sources were scheduled
but were not observed before the depletion of the instrument's
cryogenic liquid coolant.

BCDs were retrieved and the data reduced using the SSC analysis
program {\sc mopex}. We post-process individual BCD frames to remove
common MIPS artifacts (i.e., ``jail-bars'', latents etc.) and suppress
large and small-scale gradients using master-flat images generated
from the initial data. Processed frames were then background matched,
stacked, mosaicked and median filtered using {\sc mopex} to create the
final background-subtracted reduced image.

Point source extraction was performed using the SSC provided {\it
  Spitzer} Astronomical Point Source EXtraction ({\sc apex}) software
to produce $24\um$ aperture photometric fluxes of the sources in the
reduced MIPS frames (see column 5 of
Table~\ref{tab:ir_phot_spec}). The IRS spectra were convolved with the
$24 \um$ MIPS response curves and compared to the photometric fluxes
to produce absolute flux calibrated IRS spectra of each source. The
average upwards photometric correction required to the spectroscopy
was a factor of $\approx 1.3$. For the three objects lacking MIPS
observations (see Table~\ref{tab:ir_phot_spec}) we did not attempt to
correct these spectra. Hence, we consider the emission-line and
continuum fluxes (derived in
sections~\ref{subsec:spitz_emiss_line_fluxes} and
\ref{subsec:spitz_spec_decomps}) for these three AGNs to be less
accurate and most likely conservative lower-limits.

\subsection{Mid-IR Emission-line Fluxes}
\label{subsec:spitz_emiss_line_fluxes}

The reduced and flux-calibrated mid-IR spectra produced in the
previous sections were further analyzed (i.e., fitting of
emission-lines and polycyclic aromatic hydrocarbon features) using the
{\it Spitzer} spectral analysis program, {\sc smart}
\citep{smart}. Typical AGN dominated emission lines present in the
spectra of these sources included [Ne{\sc v}] ($\lambda\lambda
14.32$,$24.32\um$) and \oiv$+$[Fe{\sc ii}] ($\lambda\lambda
25.89$,$25.99\um$).\footnote{We note that due to the spectral
  resolution of the LL modules, the \oiv and [Fe{\sc ii}] emission
  lines cannot be individually resolved.}  Additionally, AGN and
star-formation produced lines such as [Ne{\sc ii}] ($\lambda
12.82\um$) and [Ne{\sc iii}] ($\lambda 15.51\um$) were also
present. See Table~\ref{tab:ir_phot_spec} for the {\it Spitzer}-IRS
derived properties and see Fig~\ref{fig:spec} for the final {\it
  Spitzer}-IRS spectra.

Mid-IR high-excitation narrow-line emission such as \nev and \oiv
(ionisation potentials of 97.1~eV and 54.9~eV, respectively) have been
shown to be excellent extinction-free unambiguous indicators of the
bolometric luminosity of an AGN (e.g.,~\citealt{melendez08b,
  diamond09, goulding10}). From analysis of our {\it Spitzer}-IRS
spectroscopy, we find that six of the AGNs in our sample have detected
\nev emission and all 14 have detected \oiv$+$[Fe{\sc
    ii}].\footnote{We note that the identification of some AGNs even
  in the very nearby Universe can often require extremely high
  signal-to-noise, high-resolution mid-IR spectroscopy (e.g.,
  \citealt{sat08, GA09}).} The \nev and \oiv luminosities for the six
AGNs with detected \nev are well correlated and lie within the
intrinsic scatter of Equation 2 of \citet{GA09} and the more recent
calibration of \citet{weaver10}; for the sources with \nev $3 \sigma$
upper-limits, we find that the majority of the fluxes are also
consistent with these relationships. Hence, we confirm that the \oiv
emission in these particular sources is likely to be a good indicator
of the intrinsic luminosity of the AGN. For AGNs with strong
contributions from star-formation, the [Fe{\sc ii}] emission may
contaminate the measured \oiv flux measured from low-resolution {\it
  Spitzer}-IRS spectroscopy \citep{melendez08a}. For those galaxies
which we find to be dominated by star formation at mid-IR wavelengths
(i.e., AGN contributions of $< 50$~percent; see
Section~\ref{subsec:spitz_spec_decomps}), we conservatively apply a
small downwards correction factor of $\approx 1.5$ \citep{melendez08a}
to our measured \oiv flux to account for the [Fe {\sc ii}]
contamination. Our final adopted \oiv luminosities cover the range,
$L_{\rm [OIV]} \approx (0.15$--$20) \times 10^{41} \ergps$.

\subsection{Spectral Decompositions}
\label{subsec:spitz_spec_decomps}
\label{subsec:6um_emission}

\begin{figure*}

\includegraphics[width=0.9\linewidth]{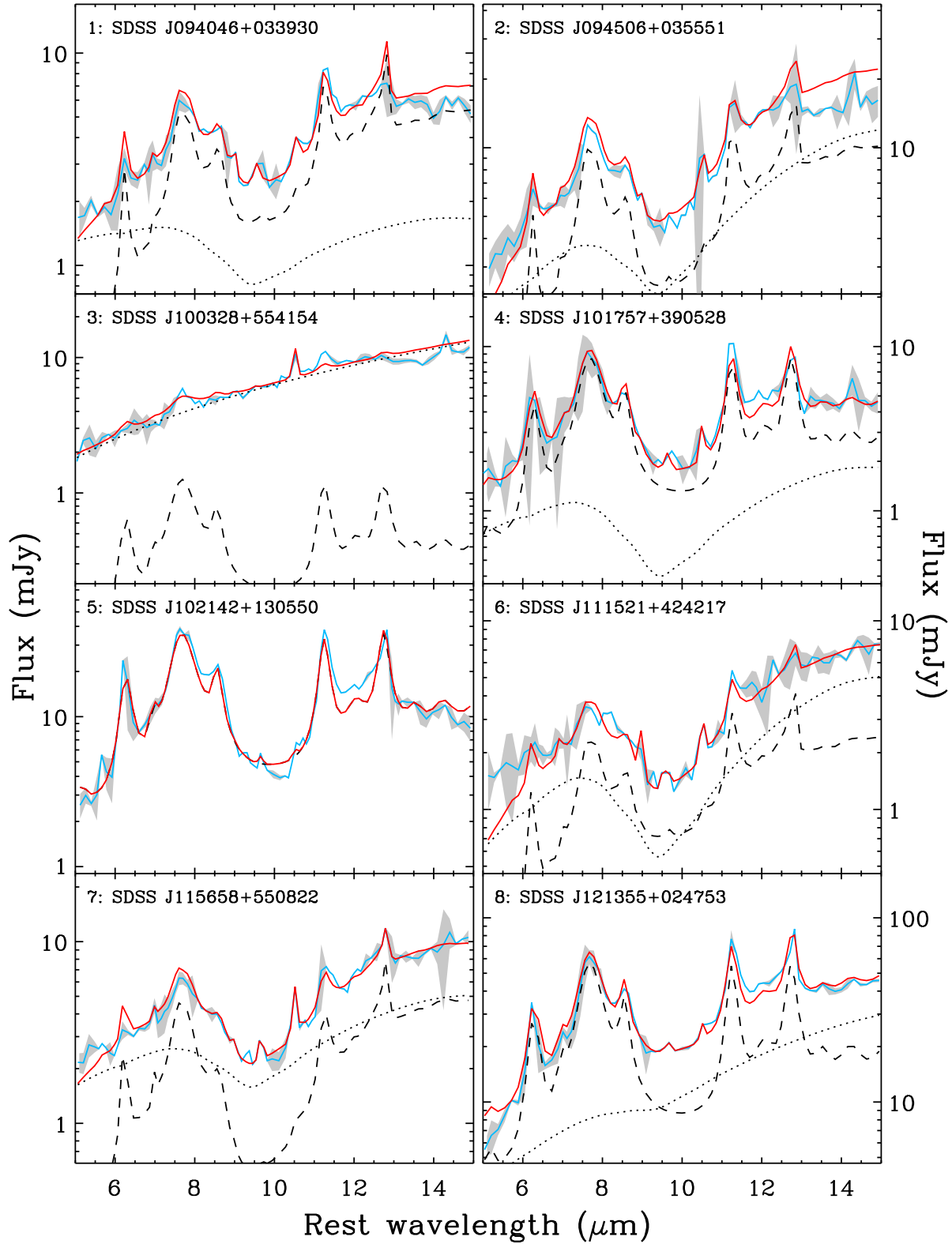}
\caption{Spectral decompositions of the {\it Spitzer}-IRS spectra
  (blue solid curve) for our candidate Compton-thick AGNs produced by
  our spectral analysis program as described in
  section~\ref{subsec:spitz_spec_decomps}. The grey shaded region
  indicates the $1\sigma$ flux uncertainties to the observed
  spectrum. The best-fit absorbed power-law and starburst template are
  shown with dotted and dashed curves, respectively. See
  Table~\ref{tab:ir_phot_spec} for best-fitting parameters. The total
  best-fit spectrum (i.e., power-law$+$starburst$+$emission-lines) is
  shown with a solid red curves.}
\label{fig:decomps}
\end{figure*}

\begin{figure*}
\hspace{1cm}
\includegraphics[width=0.9\linewidth]{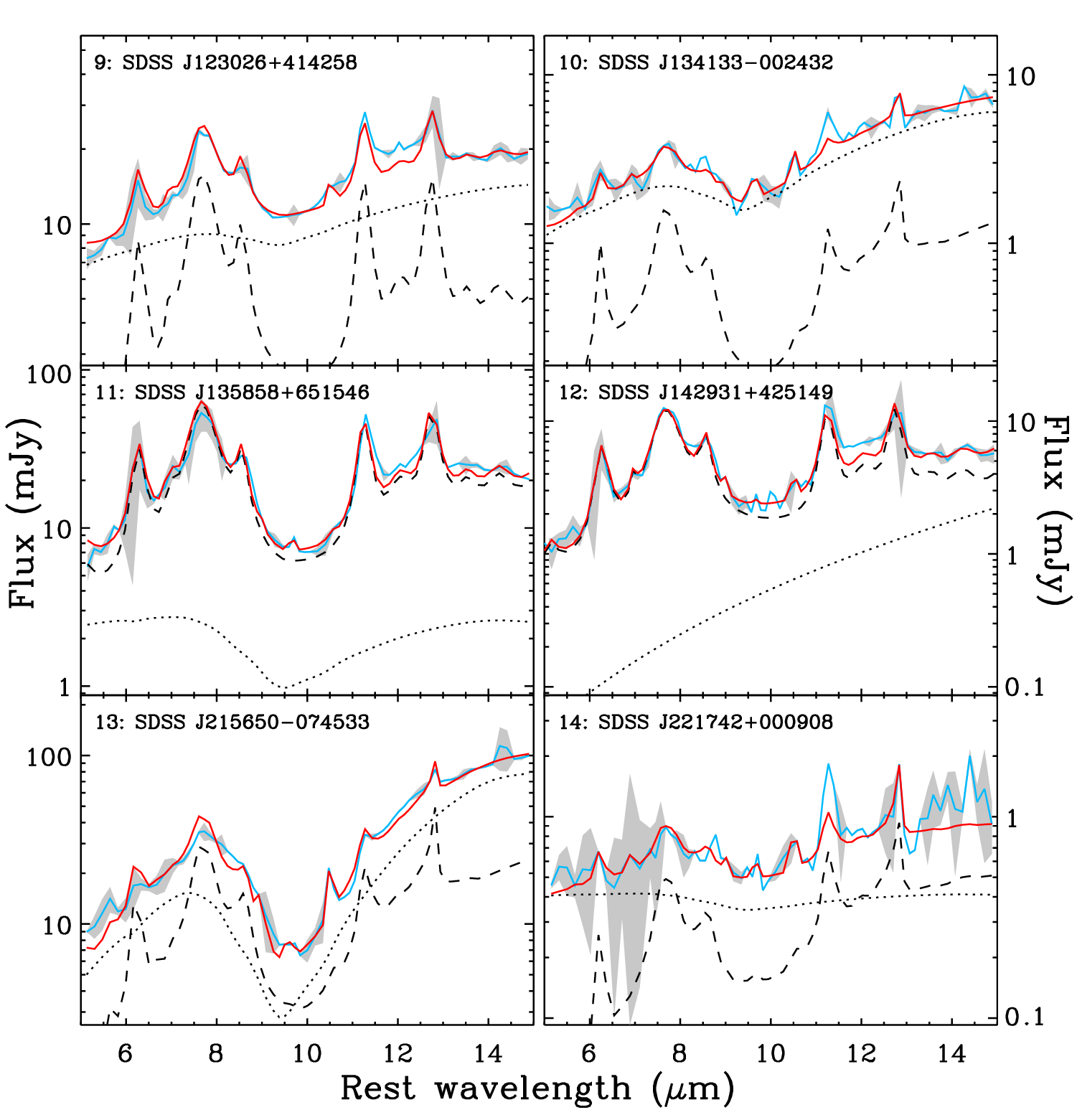}
\contcaption{}
\end{figure*}

The low-resolution mid-IR spectra of typical Type-2 AGNs at rest-frame
$\lambda \approx 4$--$15 \um$ are composed of three primary
components: 1) a power-law like thermal AGN dust continuum; 2) a
star-formation component which arises from the super-position of PAH
features; and 3) a silicate absorption feature at $\lambda \approx 9.7
\um$ produced by the hot dust continuum being absorbed by cooler dust
on parsec scales (e.g., \citealt{GA09}; \citealt{gallimore10};
\citealt{mullaney10}; \citealt{tommasin10}). Therefore, as expected,
we find that the mid-IR spectra for the majority of our candidate
Compton-thick AGNs contain both an AGN produced continuum and strong
polycyclic aromatic hydrocarbon (PAH) features, which are associated
with starburst activity in the circumnuclear photodissociation regions
of the host galaxy. Here we outline our spectral decomposition routine
to determine the relative contributions of starburst (SB) activity and
the AGN continuum in our 14 Compton-thick AGN candidates (i.e., the
SB:AGN ratio), and measure the intrinsic luminosity of the central
source from the AGN produced mid-IR continuum at $6 \um$ (e.g.,
\citealt{lutz04}).

Using a purpose-built {\sc idl}-based routine, we fit the IRS
spectroscopy for each of the 14 candidate Compton-thick AGNs with a
combined standard starburst template and an AGN power-law component
(with spectral index, $k$) convolved with a \citet{draine07}
extinction curve ($\rho ( \lambda )$) of the form,
\begin{equation}
  f_{AGN}(\lambda) = a \lambda^{k} {\rm exp}[-b \tau \rho ( \lambda )]
\end{equation}
where $a$, $b$ and $k$ are constants, and $\tau$ is the optical
depth.\footnote{Our {\sc idl} routine is based on that used by
  Mullaney et~al. (2010b) and makes use of the Markwardt 1-dimensional
  Chi-squared analysis library, see
  http://cow.physics.wisc.edu/$\sim$craigm/idl/ for further details.}
Within the fitting we use four possible starburst templates which
cover a realistic range of physical and theoretical scenarios: 1)
low-resolution {\it Spitzer}-IRS spectroscopy of the archetypal
starburst galaxy, M82; 2) a combined {\it Spitzer}-IRS starburst
template of local pure star-forming galaxies presented in
\citet{brandl06};\footnote{We note that as we require only starburst
  emission in these templates, we do not include galaxies in the
  combined Brandl starburst template which have any previous evidence
  for AGN activity (i.e., Mrk~266, NGC~660, 1097, 1365, 3628 and
  4945).} 3) a theoretical radiative transfer model of a pure
circumnuclear starburst region at $r \approx 3$~kpc with $L_{\rm IR}
\approx 10^{10} \Lsun$ (\citealt{siebenmorgen07}; hereafter, SK07);
and 4) a theoretical radiative transfer model of a nuclear star
cluster at $r < 0.35$~kpc with $L_{\rm IR} \approx 10^{10} \Lsun$
(SK07).

The best resulting model parameters derived from the minimum
Chi-squared fit to the IRS data are given in
Table~\ref{tab:ir_phot_spec} and shown in Fig.~\ref{fig:decomps}. We
note that none of the AGNs in our sample have mid-IR spectral features
which are consistent with the theoretical nuclear star cluster model,
and hence, the best-fit spectral model for each are that of an AGN
combined with one of the three circumnuclear starburst templates.
Based on the mid-IR spectral-fits, we also derive the approximate
contribution of the AGN to the mid-IR emission for each of the
sources; see column (7) of Table~\ref{tab:ir_phot_spec}. We find that
although these AGNs were selected to be strong \oiii and [N{\sc ii}]
emitters (i.e., optically-dominated Seyfert galaxies), the mid-IR
spectra of $\approx 50$ percent of the sources are consistent with
being dominated by star-formation activity. Indeed, on the basis of
these spectral decompositions, the mid-IR spectroscopy for one source
(SDSS~J102142$+$130550) is consistent with there being no AGN
component, despite this source clearly being identified as an AGN at
optical wavelengths.

\begin{table*}
\begin{minipage}{175mm}
\begin{center}
\setlength{\tabcolsep}{0.25mm}
\caption{Measured AGN properties}
\label{tab:ir_phot_spec}
\begin{tabular}{lcccccccccccc}
\hline
  \multicolumn{1}{|c|}{ID} &
  \multicolumn{1}{|c|}{\nev$\lambda 14.32$} &
  \multicolumn{1}{|c|}{\nev$\lambda 24.32$} &
  \multicolumn{1}{|c|}{\oiv$\lambda 25.89$} &
  \multicolumn{1}{|c|}{$S_{\rm 24 \mu m}$} &
  \multicolumn{1}{|c|}{SB} &
  \multicolumn{1}{|c|}{AGN} &
  \multicolumn{1}{|c|}{$S_{\rm 6 \mu m}$} &
  \multicolumn{1}{|c|}{log($L_{\rm X,[OIV]}$)} &
  \multicolumn{1}{|c|}{log($L_{\rm X,6 \mu m}$)} &
  \multicolumn{1}{|c|}{log($\eta$)} &
  \multicolumn{2}{|c|}{C-thick?} \\
  \multicolumn{1}{|c|}{} &
  \multicolumn{1}{|c|}{(erg s$^{-1}$ cm$^{-2}$)} &
  \multicolumn{1}{|c|}{(erg s$^{-1}$ cm$^{-2}$)} &
  \multicolumn{1}{|c|}{(erg s$^{-1}$ cm$^{-2}$)} &
  \multicolumn{1}{|c|}{(mJy)} &
  \multicolumn{1}{|c|}{Model} &
  \multicolumn{1}{|c|}{cont.} &
  \multicolumn{1}{|c|}{(mJy)} &
  \multicolumn{1}{|c|}{($\ergps$)} &
  \multicolumn{1}{|c|}{($\ergps$)} &
  \multicolumn{1}{|c|}{} &
  \multicolumn{1}{|c|}{\oiv} &
  \multicolumn{1}{|c|}{$6 \um$}  \\
  \multicolumn{1}{|c|}{(1)} &
  \multicolumn{1}{|c|}{(2)} &
  \multicolumn{1}{|c|}{(3)} &
  \multicolumn{1}{|c|}{(4)} &
  \multicolumn{1}{|c|}{(5)} &
  \multicolumn{1}{|c|}{(6)} &
  \multicolumn{1}{|c|}{(7)} &
  \multicolumn{1}{|c|}{(8)} &
  \multicolumn{1}{|c|}{(9)} &
  \multicolumn{1}{|c|}{(10)} &
  \multicolumn{1}{|c|}{(11)} &
  \multicolumn{1}{|c|}{(12)} &
  \multicolumn{1}{|c|}{(13)} \\
\hline
1  & $<0.98$           & $<0.77$          & $3.91   \pm 0.81$ & $3.98 \pm 0.04$   & 2 & 40\%     & $1.71 \pm 0.14$  & 42.67 & 42.74 & -2.24 & y & y \\
2  & $10.51 \pm 1.43$  & $6.60 \pm 2.09$  & $29.39  \pm 1.05$ & $17.82 \pm 0.05$  & 2 & 55\%     & $3.27 \pm 0.19$  & 44.42 & 43.58 & -0.94 & Y & y \\
3  & $4.07 \pm 0.78$   & $4.07 \pm 0.78$  & $6.45   \pm 2.03$ & $8.49 \pm 0.03$   & 3 & 92\%     & $2.49 \pm 0.02$  & 43.63 & 43.30 & -1.02 & y & ? \\
4  & $3.58 \pm 0.55$   & $3.65 \pm 0.54$  & $12.89  \pm 0.64$ & -                 & 3 & 38\%     & $1.30 \pm 0.76$  & 42.76 & 42.08 & -2.28 & Y & ? \\
5  & $<2.94$           & $<1.00$          & $6.79   \pm 1.62$ & $8.90 \pm 0.05$   & 3 & $<0.1$\% & $<0.01$          & 42.80 & -     & -     & Y & - \\
6  & $<1.67$           & $<0.78$          & $5.64   \pm 0.91$ & $6.80 \pm 0.03$   & 2 & 72\%     & $1.42 \pm 0.17$  & 43.88 & 43.34 & -     & Y & y \\
7  & $3.33 \pm 0.98$   & $2.95 \pm 0.95$  & $10.90  \pm 1.38$ & $10.72 \pm 0.10$  & 2 & 53\%     & $2.53 \pm 0.31$  & 43.26 & 42.74 & -1.92 & Y & y \\
8  & $7.40 \pm 1.81$   & $<2.19$          & $19.07  \pm 1.03$ & $65.36 \pm 0.06$  & 3 & 38\%     & $5.47 \pm 0.99$  & 43.26 & 43.01 & -0.63 & Y & Y \\
9  & $<2.07$           & $<1.47$          & $5.24   \pm 0.90$ & -                 & 3 & 66\%     & $8.36 \pm 0.33$  & 43.45 & 43.69 & -0.50 & Y & Y \\
10 & $<1.08$           & $<1.32$          & $6.52   \pm 0.92$ & $6.93 \pm 0.04$   & 1 & 85\%     & $1.95 \pm 0.11$  & 42.91 & 42.52 & -2.75 & y & ? \\
11 & $<2.46$           & $<3.39$          & $12.89  \pm 1.92$ & $39.94 \pm 0.03$  & 3 & 12\%     & $3.47 \pm 2.71$  & 42.26 & 42.06 & -1.87 & y & y \\
12 & $<1.07$           & $<0.63$          & $4.68   \pm 0.76$ & $6.33 \pm 0.03$   & 3 & 9\%      & $0.09 \pm 0.40$  & 43.35 & 41.91 & -2.79 & ? & ? \\
13 & $57.16 \pm 3.16$  & $42.17 \pm 6.52$ & $180.77 \pm 6.01$ & $106.90 \pm 0.55$ & 1 & 81\%     & $16.42 \pm 3.20$ & 44.24 & 43.23 & -0.73 & Y & Y \\
14 & $<1.44$           & $<0.96$          & $3.06   \pm 0.51$ & -                 & 2 & 62\%     & $0.43 \pm 0.09$  & 42.09 & 41.44 & -2.10 & Y & ? \\

  \hline\end{tabular}
\medskip
\end{center}

{\tiny NOTES:} (1) SDSS source-identification. (2-4) Fluxes and their
$1 \sigma$ uncertainty for the measured AGN-produced mid-IR emission
lines in units of $10^{-15} $~erg s$^{-1}$ cm$^{-2}$. (5) MIPS $24
\um$ flux density in units of mJy. (6) Best-fit starburst model in the
mid-IR spectral modeling analysis (see
section~\ref{subsec:spitz_spec_decomps}) - 1:~M82; 2:~Combined SB
(Brandl et~al. 2006); 3:~Circumnuclear starburst at 3~kpc (SK07). (7)
Inferred AGN contribution to the mid-IR emission at $\lambda \sim
5$--$15 \um$. (8) Unabsorbed AGN $6 \um$ flux and $1 \sigma$
uncertainty in units of mJy. (9-10) Logarithm of the estimated
2--10~keV luminosity from the AGN-produced \oiv and $6 \um$ emission
(see sections~\ref{subsec:spitz_emiss_line_fluxes},
\ref{subsec:6um_emission} and \ref{subsec:intrinsic_nh_from_oiv} for
further details). (11) Logarithm of implied Eddington ratio derived
from the estimated 2--10~keV luminosity using $6 \um$ emission, and
the SMBH masses presented in Table~1. (12-13) Indication for whether
the source is consistent with being Compton-thick ($N_H > 1.5 \times
10^{24} \pcmsq$) on the basis of the mid-IR indicators: `Y' -
conservatively identified to be Compton-thick, `y' -
less-conservatively identified to be Compton-thick, `?' - inconclusive
due to depth of current X-ray data (i.e., they are upper-limits; see
section~\ref{subsec:intrinsic_nh_from_oiv}).

\end{minipage}
\end{table*}

In order to estimate the intrinsic AGN luminosity and hence place
limits on the X-ray absorption in these sources, we use the measured
AGN power-law parameters to derive $6\um$ luminosities ($L_{6 \mu
  m}$). The uncertainties of these $6 \um$ fluxes are established by
considering a weighted spread in the measured $6 \um$ fluxes from all
statistically valid starburst template fits (i.e., we reject all
statistically poor fits at the 95 per cent level). The mean $1\sigma$
uncertainty is $\approx 0.1$~dex. See column (8) of
Table~\ref{tab:ir_phot_spec}.

We estimate the $6 \um$ continuum luminosities for 13 of our 14
candidate Compton-thick AGNs and conservatively estimate an upper
limit for the $6 \um$ continuum flux of $< 10^{-2}$~mJy for
SDSS~J102142$+$130550. We find using our adopted cosmology, that the
AGNs cover more than 2 decades in $6\um$ luminosity, with $\nu L_{6\mu
  m} \approx (0.1$--$20) \times 10^{43} \ergps$.


\section{Results and Discussion}
\label{sec:g10b_results}

We have selected a sample of 14 \oiii bright, X-ray undetected AGNs
from the $\approx 100$~deg$^2$ overlap region between the SDSS-DR7 and
2XMMi surveys. These sources lie at $z \sim 0.03$--0.2 and host
moderate to highly luminous AGNs with de-reddened $L_{\rm [OIII]}
\approx (0.2$--$500) \times 10^{42} \ergps$ (i.e., similar to those of
typical nearby Seyfert galaxies). Our 14 targets all have $f_X /
f_{\rm [OIII]} < 1$, implying strong intrinsic absorption of their
X-ray flux and many (possibly all) are likely Compton-thick AGNs. In
the absence of X-ray spectroscopic data, in
section~\ref{sec:midir_spec_photom} we derived AGN-produced emission
line and continuum luminosity measurements in order to independently
constrain the intrinsic luminosity of these candidate Compton-thick
AGNs. In this section, we use these intrinsic luminosities in
conjunction with X-ray constraints from {\it XMM-Newton} data to test
whether these objects are indeed Compton-thick AGNs. We then use these
results to place new constraints on the space density and relative
mass-accretion rates of Compton-thick AGNs in the nearby Universe ($z
\sim 0.1$).

\subsection{Identifying Compton-thick AGNs at $z \sim 0.1$}
\label{subsec:intrinsic_nh_from_oiv}
\label{subsec_intrinsic_nh_from_6um}

Previously, strong, relatively high-excitation emission lines, such as
\oiii$\lambda 5007$ (35.1~eV), have been used as proxies for hard
X-ray emission in AGNs, and hence, their intrinsic luminosity ($L_{\rm
  AGN}$; e.g.,
\citealt{mulchaey94,alonso97,heckman05,panessa06}). However, such
emission may also be readily excited by strong star formation as well
as being subject to significant dust extinction within the host
galaxy. By contrast, mid-IR high-excitation narrow-line emission
(e.g., [Ne{\sc v}]; [O{\sc iv}]) is an excellent extinction-free
indicator of $L_{\rm AGN}$ (see
Section~\ref{subsec:spitz_emiss_line_fluxes}) and, when combined with
sensitive X-ray data, can provide good first order constraints on
whether an AGN is Compton thick.

\begin{figure*}
\hspace{-1.cm}
\includegraphics[width=0.54\textwidth]{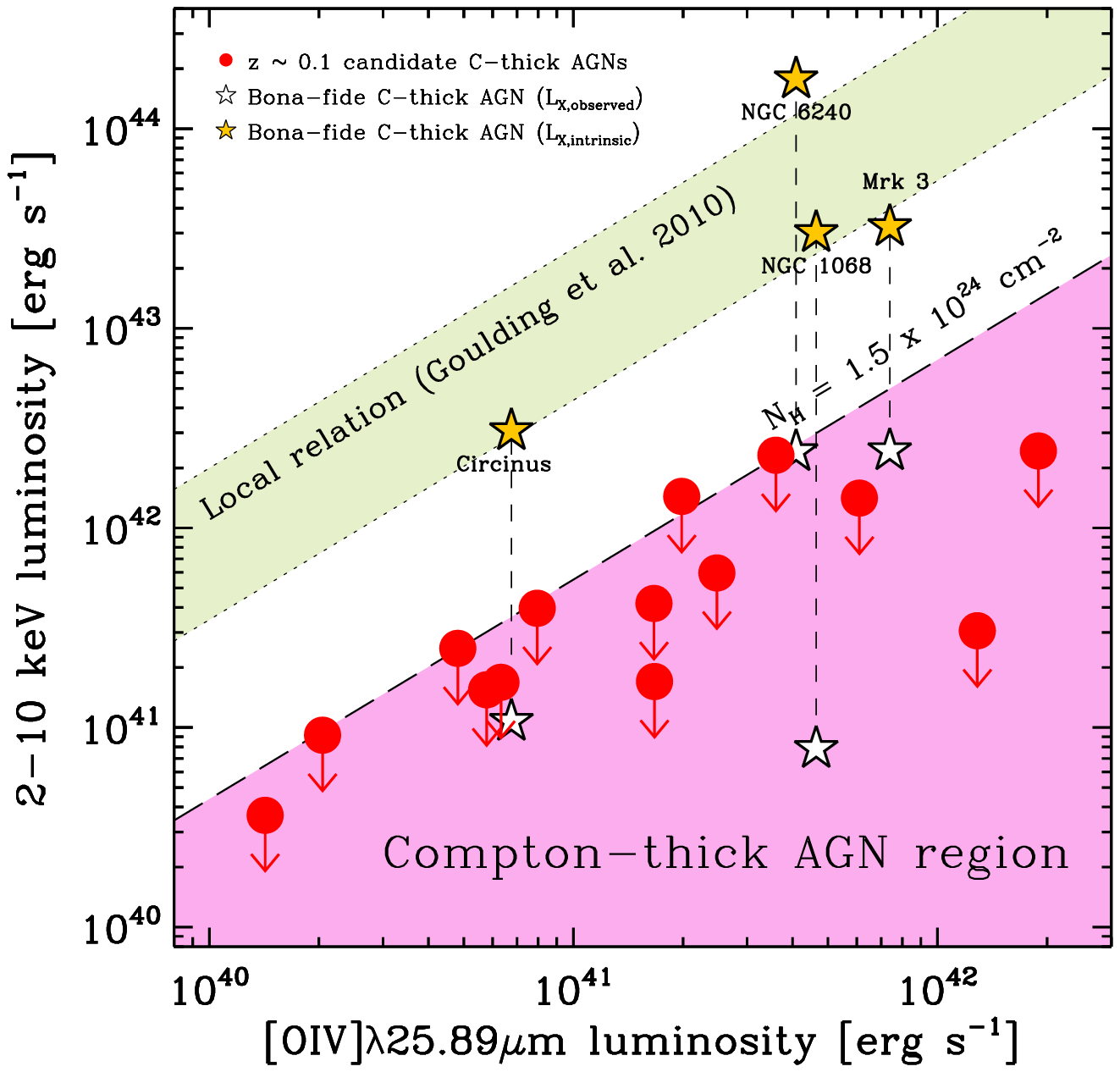}
\hspace{-1.cm}
\includegraphics[width=0.54\textwidth]{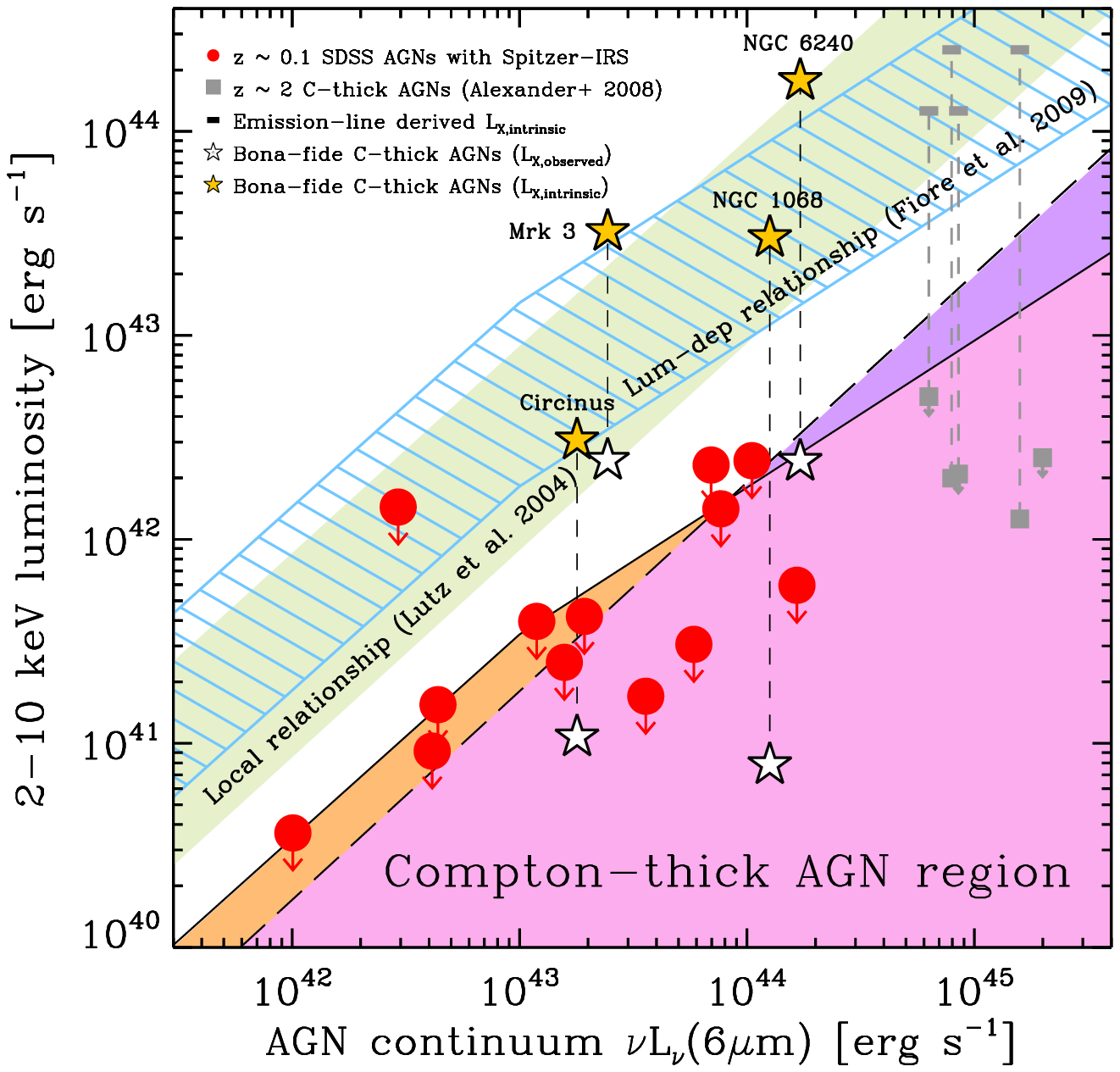}
\caption{{\bf Left; (a):} Rest-frame 2--10 keV X-ray luminosity versus
  mid-IR \oiv ($\lambda 25.89 \um$) luminosity ($L_{\rm [OIV]}$) for
  our sample of Compton-thick AGN candidates (filled circles). We show
  the local X-ray--\oiv relation of Goulding et al. (2010) and find on
  the basis of $L_{\rm [OIV]}$ that the majority of our sample are
  consistent with the observed X-ray luminosity being absorbed by a
  factor $\geq 15$ (i.e., $N_H > 1.5 \times 10^{24} \pcmsq$). {\bf
    Right (b):} Rest-frame 2--10~keV X-ray luminosity versus the
  mid-IR AGN continuum luminosity at $6 \um$ for the Compton-thick AGN
  candidates with measured $6 \um$ luminosities (filled circles). A
  comparison sample of $z \sim 2$ Compton-thick quasars is
  additionally shown (Alexander et~al. 2008; grey squares). We use the
  local X-ray--$6\um$ relation of Lutz et~al. (2004) and the
  luminosity-dependent relation of Fiore et~al. (2009) to predict the
  region of parameter space where Compton-thick AGN lie. We find that
  on the basis of $6 \um$ luminosities, many of the sources in our
  sample are likely to be Compton-thick AGNs. For comparison, in both
  (a) and (b) we additionally highlight the observed (filled stars)
  and absorption-corrected (open stars) X-ray luminosities for 4
  well-studied `bona-fide' Compton-thick AGNs (Circinus, Mrk~3,
  NGC~1068 and 6240). In both figures, the observed X-ray luminosities
  for these `bona-fide' Compton-thick AGNs occupy roughly the same
  region of parameter space as our sample but our objects are $\approx
  10$--100 times more distant. However, their intrinsic X-ray
  luminosities are consistent with the local relations.}
\label{fig:6um_xray}
\end{figure*}

In Fig.~\ref{fig:6um_xray}a we present the observed 2--10~keV X-ray
upper-limit luminosities from the {\it XMM-Newton} data versus the
mid-IR \oiv luminosity for our candidate Compton-thick AGNs and
compare them to the intrinsic properties found for local `bona-fide'
Compton-thick AGNs. We find that the candidate Compton-thick AGNs are
spread over a wide range of \oiv luminosities, $L_{\rm [OIV]} \approx
(0.13$--$20) \times 10^{41} \ergps$; for the sources in our sample
which we find to be dominated by SF at mid-IR wavelengths (column 7 of
Table~\ref{tab:ir_phot_spec}), \oiv fluxes have conservatively been
adjusted for contamination from [Fe{\sc ii}] emission (see
section~\ref{subsec:spitz_emiss_line_fluxes}).  We find that based on
the observed X-ray upper-limits, none of the objects in our sample are
consistent with the local intrinsic relation of AGNs from
\citet{goulding10}, suggesting that the X-ray emission is heavily
obscured. However, as we illustrate in Fig.~\ref{fig:6um_xray}a, the
observed $L_X / L_{\rm [OIV]}$ for our sample (mean ratio $\approx
3.6$) is consistent with the observed $L_X / L_{\rm [OIV]}$ ratio for
a sample of well-studied local `bona-fide' Compton-thick AGNs (i.e.,
Circinus, Mrk~3, NGC~1068 and NGC~6240). Furthermore, the $L_X /
L_{\rm [OIV]}$ luminosity ratio of these four Compton-thick AGNs is
consistent with the $L_X $--$ L_{\rm [OIV]}$ relationship of Goulding
et~al. (2010) when the X-ray data is corrected for the absorption
implied from high-quality X-ray spectroscopy. Assuming the \oiv
emission is indeed an isotropic AGN indicator (e.g., Melendez
et~al. 2008; Diamond-Stanic et~al. 2009; Goulding et~al. 2010), this
suggests that by comparing the observed X-ray upper-limit to the
intrinsic X-ray luminosity as predicted by our \oiv measurements
($L_{\rm x,[OIV]}$), we may infer whether the sources in our sample
are indeed Compton-thick AGNs.

Based on Compton reflection models, Alexander et~al. (2008) predict
that the observed--intrinsic X-ray flux ratio in the 2--10~keV band
for a Compton-thick AGN with $N_H \sim 1.5 \times 10^{24} \pcmsq$ is
$f_{\rm X,intr}/f_{\rm X,obs} \approx 15$. Hence, we predict that
sources with $f_{\rm X,intr}/f_{\rm X,obs} \goa 15$ are likely to be
obscured by Compton-thick material. For the 14 candidate Compton-thick
AGNs in our sample, we calculate $f_{\rm X,[OIV]}$ using the local
X-ray--\oiv relation of \citet{goulding10}. We predict intrinsic X-ray
luminosities of $L_{X{\rm ,predict}} \approx (0.1$--$26) \times
10^{43} \ergps$ (see Column 9 of Table~\ref{tab:ir_phot_spec}). We
find that 13 ($\approx 90$~percent) of the sources exhibit $f_{\rm
  X,[OIV]}/f_{\rm X,obs} \goa 15$ (see Column 12 of
Table~\ref{tab:ir_phot_spec}). Furthermore, if we account for the
intrinsic scatter within the local relation ($\approx 0.3$~dex), and
conservatively assume that none of the sources which lie within this
region are Compton-thick, we still estimate that at least 9/14
($\approx 65$~percent) of our candidate Compton-thick AGNs could be
genuine Compton-thick AGNs. It is also prudent to note that as the
observed X-ray fluxes for all of these sources are upper-limits, we
cannot exclude the possibility that all of the sources in our sample
are Compton-thick AGNs as the implied $f_{\rm X,[OIV]}/f_{\rm X,obs}$
ratio is a lower-limit.

By combining multiple indirect AGN luminosity indicators, particularly
those which probe different regions of the central engine, we can
place even stronger constraints on whether the AGNs in our sample are
Compton thick than using narrow-line emission alone. The $6 \um$
continuum luminosity has been shown to provide a good proxy for the
intrinsic AGN luminosity (e.g., Lutz et~al. 2004;
\citealt{Maiolino07,Treister08,fiore09}). In Fig.~\ref{fig:6um_xray}b
we again present the observed 2--10~keV X-ray upper-limit luminosities
from {\it XMM-Newton} data but now compare these luminosities to the
AGN continuum luminosity at $6 \um$ derived in
Section~\ref{subsec:6um_emission} and the luminosity-dependent
\citep{fiore09} and luminosity-independent (Lutz et~al. 2004)
relations derived using high-quality X-ray data and mid-IR {\it
  Spitzer} IRAC photometry and ISO spectroscopy, respectively. As
noted in Section~\ref{subsec:6um_emission}, one of our 14 candidate
Compton-thick AGNs is consistent with there being little or no mid-IR
emission from an AGN continuum at $\lambda \sim 5$--$15 \um$, and we
remove this AGN from further analyses in here.

We conservatively adopt the slightly lower-luminosity X-ray--$6 \um$
relationship of Lutz et~al (2004) to infer the intrinsic X-ray
luminosities of the candidate Compton-thick AGNs. We estimate
intrinsic X-ray luminosities of $L_{X {\rm ,predict}} \approx
(0.2$--$30) \times 10^{42} \ergps$ (see Column 10 of
Table~\ref{tab:ir_phot_spec}). Eight out of the 13 ($\approx
60$~percent) $6 \um$ detected sample members lie in the region
expected for Compton-thick AGNs (i.e., $N_H \goa 1.5 \times 10^{24}
\pcmsq$; see Column 13 of Table~\ref{tab:ir_phot_spec}). However, if
we were to adopt the relationship of Fiore et~al. (2009) this
Compton-thick AGN fraction would increase to 9/13 sources ($\approx
80$~percent; i.e., consistent with that found when using \oiv as a
$N_H$ diagnostic). Furthermore, if we account for the intrinsic
scatter of within the Lutz et~al (2004) relationship ($\approx
0.5$~dex), and (as above) conservatively assume that none of the
sources which lie within this region of scatter are genuine
Compton-thick AGNs, we still find that at least 3 of the 13 ($\approx
20$~percent) sources must be Compton thick.

In Columns 12 and 13 of Table~\ref{tab:ir_phot_spec} we summarise
whether we identify the sources to be Compton-thick AGNs on the basis
of their combined X-ray and mid-IR properties. We consider those AGNs
which are conservatively identified as Compton-thick AGNs (i.e., those
which lie below the region of intrinsic scatter derived from the
$L_X$--$L_{\rm [OIV]}$ and $L_X$--$L_{6 \mu m}$ relationships) in at
least one of the mid-IR diagnostics and are also below the standard
Compton thick threshold in the other mid-IR diagnostic to be genuine
Compton-thick AGNs (i.e., in the nomenclature of
Table~\ref{tab:ir_phot_spec}, only those AGNs with Y-Y, Y-y or
y-Y). This is a reasonable and conservative assumption to make if we
consider the `bona-fide' Compton-thick AGNs shown in
Fig.~\ref{fig:6um_xray}; all of these AGNs would be identified to be
Compton-thick AGNs (y/Y) on the basis of \oiv emission and 3/4 on the
basis of $6 \um$ emission. Hence, using our adopted definition, we
find that 6/14 ($\approx 43$~percent) of our candidate Compton-thick
AGNs are very likely genuine Compton-thick AGNs on the basis of their
combined mid-IR properties. Under the reasonable assumption that our
sample of candidate Compton-thick AGNs is a representative sub-sample
of the parent X-ray undetected population of AGNs in the SDSS-DR7
(i.e., in both redshift and luminosity parameter space; see
section~\ref{sec:g10b_sampleselect}), these results imply that at
least $\approx 43 \pm 21$ percent of the sources with $f_X / f_{\rm
  [OIII]} < 1$ are Compton thick.\footnote{Uncertainties are
  calculated using standard Poisson counting statistics.} We note that
on the basis of our most conservative and most optimistic
Compton-thick diagnostic thresholds, $\approx 20$--100 percent of AGNs
with $f_X / f_{\rm [OIII]} < 1$ could be Compton thick.

Of the four AGNs which were selected because they lie in the
exclusively Compton-thick AGN region of Fig.~\ref{fig:XO3} (i.e.,
those with $f_X / f_{\rm [OIII]} < 0.1$), three have mid-IR
emission-line and continuum AGN indicators consistent with the X-ray
emission being absorbed by at least a factor $\goa 15$. Hence, these
are very likely to be Compton-thick AGNs. Whilst the \oiv emission
from the fourth AGN (SDSS~J221742$+$000908) is consistent with a
Compton-thick AGN (see Fig.~\ref{fig:6um_xray}a) we find little
evidence for this on the basis of its $6 \um$ continuum
luminosity. Indeed, the observed $L_{X,2-10keV}$ appears to be
comparatively unabsorbed on the basis of AGN continuum
luminosity. However, we note that we would also find a similar result
for the `bona-fide' Compton-thick AGN, Mrk~3. We find that five of the
10 AGNs with $f_X / f_{\rm [OIII]} \sim 0.1$--1.0 exhibit mid-IR
emission features consistent with Compton-thick AGNs. All five AGNs
have strong \oiv and $6 \um$ luminosities suggesting strong absorption
of the X-ray emission, as well as evidence for silicate absorption at
$9.7 \um$. By contrast, for the AGN which does not appear to be
clearly Compton-thick on the basis of either of our mid-IR AGN
indicators (SDSS~J142931$+$425149), we find that the underlying AGN
continuum in this sources is consistent with an unabsorbed power-law
(i.e., no evidence for silicate absorption).

\subsection{The space-density of Compton-thick AGNs at $z \sim 0.1$}
\label{subsec_space_dens_cthick_agn}

Based on the mid-IR emission-line and continuum emission diagnostics,
we find that at least six ($\goa 43 \pm 21$~percent) of our sample of
14 X-ray undetected optical narrow-line AGNs with $L_{X}/L_{\rm
  [OIII]} < 1$ appear to suffer from heavy intrinsic absorption with
$N_H \goa 1.5 \times 10^{24} \pcmsq$ (i.e., they are Compton-thick
AGNs). Assuming that our sample of candidate Compton-thick AGNs is
representative of the parent population, we may use our derived
Compton-thick AGN fraction to infer at least a lower limit for the
number, and hence space density, of Compton-thick AGNs at $z \sim
0.03$--0.2.

\begin{figure}
\includegraphics[width=1.00\linewidth]{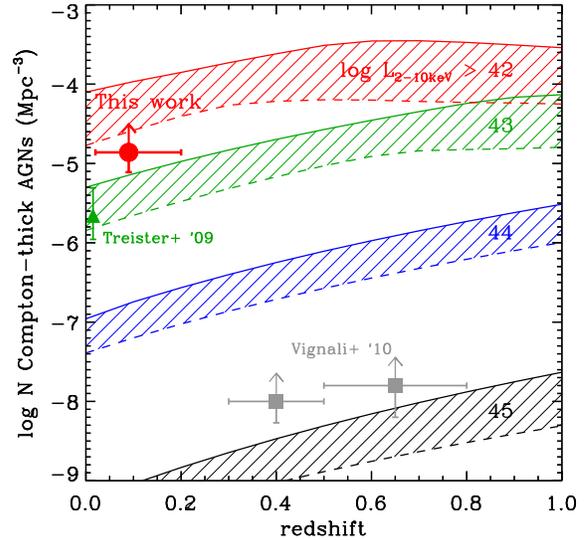}
\caption{Space density of Compton-thick AGNs compared with the XRB
  synthesis models of Gilli et~al. (2007; solid curves) and Treister
  et~al. (2009; dashed curves) for intrinsic X-ray luminosities of
  $L_{\rm X} > 10^{42}$,$10^{43}$,$10^{44}$ and $10^{45} \ergps$. Data
  points refer to this work (solid circle; $L_{\rm X} > 10^{42}
  \ergps$) and those from comparable studies of $z \sim 0$
  Compton-thick AGNs (Treister et~al. 2009; solid triangle; $L_{\rm X}
  > 10^{43} \ergps$) and luminous Compton-thick quasars at $z \sim
  0.3$--0.8 derived from the SDSS (Vignali et~al. 2010; solid squares;
  $L_{\rm X} > 2.5 \times 10^{44} \ergps$).}
\label{fig:space_dens}
\end{figure}

The total number of Type-2 AGNs in the SDSS-2XMMi overlap region is
334 at $z \sim 0.03$--0.2, 147 ($\approx 45$~percent) of these AGNs
are X-ray undetected in 2XMMi with $L_{X}/L_{\rm [OIII]} < 1$ (i.e.,
our parent sample). Based on our derived Compton-thick AGN fraction
above, we would expect that $\approx 63 \pm 31$ of the sources in our
parent sample with $L_{X}/L_{\rm [OIII]} < 1$ to be Compton-thick
AGNs, i.e., $\approx 20$~percent of all Type-2 AGNs in the SDSS-2XMMi
overlap region. The comoving volume encompassed by our survey is $V
\approx 4.6 \times 10^6$~Mpc$^3$. Hence, we estimate a space-density
of Compton-thick AGNs at $z \sim 0.1$ of log$(\Phi) \approx
-4.9^{+0.2}_{-0.3}$~Mpc$^{-3}$ for luminosities of $L_X \goa 10^{42}
\ergps$. However, given that both our sample selection and method for
identifying Compton-thick AGNs are based on X-ray flux upper-limits
(i.e., those sources with $L_{X}/L_{\rm [OIII]} > 1$ which currently
lack formal X-ray detections may have $L_{X}/L_{\rm [OIII]} < 1$ with
deeper X-ray observations; all of the AGNs in our {\it Spitzer}-IRS
sample may still be Compton thick), the space-density found here
should be considered a strict lower-limit to the number of
Compton-thick AGNs in our parent sample.

In Fig.~\ref{fig:space_dens} we compare our estimated space-density to
the similar study of SDSS-selected Compton-thick quasars of
\citet{vignali10} and to those derived from the XRB synthesis models
of Gilli et~al. (2007) and Treister et~al. (2009).\footnote{The XRB
  synthesis models of Gilli et~al. (2007) and Treister et~al. (2009)
  are defined using a slightly different cosmology ($H_0 = 70 \kmps$)
  to that considered here ($H_0 = 71 \kmps$). However, this has little
  or no effect on the conclusions drawn from the comparisons.} Both
models invoke a large population of Compton-thick AGNs in order to fit
the peak of the XRB at $E \sim 30$~keV. The differences in the
required space densities of Compton-thick AGNs are derived from the
normalisations to the XRB, as well as the inclusion of the local
Compton-thick AGNs identified by {\it INTEGRAL} and {\it Swift} (green
triangle in Fig.~\ref{fig:space_dens}) in the model of Treister
et~al. (2009). We find that at the median redshift of our parent
sample, our estimated space density is systematically lower than those
predicted by the Gilli et~al. (2007) and Treister et~al. (2009) models
by a factor of $\approx 8$ and 2, respectively for $L_X \goa 10^{42}
\ergps$. It is important to note that our estimated space density is a
lower limit, and is therefore consistent with both model
predictions. However, if we were to assume the highly unlikely
situation that all of our parent sample of 334 AGNs are Compton thick
(i.e., a predicted space density of log$(\Phi) \approx -4.14$), this
would still be systematically below the predictions of Gilli
et~al. (2007).

The discrepancy between our observations and the model expectations
can be explained in several ways. The most obvious explanation is that
our parent sample of optical SDSS NL-AGNs does not include all
obscured AGNs at $0.03 < z < 0.2$. In the SDSS-DR7, only the most
luminous galaxies are targeted for spectroscopic follow-up, resulting
in very few ($\loa 10$~percent) galaxies being included in the SDSS
with $M_* \loa 10^{10} \Msun$ (i.e., those likely to host relatively
low mass SMBHs with $\Mbh \loa 10^7 \Msun$) to $z \sim 0.1$ (e.g.,
\citealt{kauff03a}). Hence, our parent sample may not include the
majority of AGNs with relatively low mass SMBHs. Using our
Compton-thick AGN sample we can estimate the number of low mass SMBHs
not included in our derived space density. Given the $L_{\rm X} /
\Mbh$ distribution of our sample (SMBH masses for our sample are
estimated from stellar velocity dispersions; see Section 4.3 for
further details; see Column 7 of Table~1), $\approx (0.2$--$20) \times
10^{35} \ergps \Msun^{-1}$, at the luminosity completeness limit
($L_{\rm X} \goa 10^{42} \ergps$) and assuming a similar distribution
in Eddington ratios, the smallest SMBH which could conceivably be in
our sample is $\Mbh \approx 5 \times 10^5 \Msun$; this is a factor
$\approx 10$ below the lowest mass Compton-thick AGN identified here
($\Mbh \approx 5 \times 10^6 \Msun$). We may now estimate how many
Compton-thick AGNs may contain SMBHs in the mass region $\Mbh \approx
(0.5$--$5) \times 10^6 \Msun$, and hence constrain the number of
Compton-thick AGNs not included in our space-density estimate due to
the lower mass limit of the SDSS.

If the five most conservatively identified Compton-thick AGNs with
$\Mbh$ estimates contained SMBHs which were a factor $\approx 10$
smaller in mass but had the same $L_{\rm X} / \Mbh$ ratio, four
($\approx 80$~percent) would still have $L_{\rm X} \goa 10^{42}
\ergps$ and would therefore be included in our estimate of the space
density of Compton-thick AGNs as shown in
Fig.~\ref{fig:space_dens}. Based on a simple extrapolation of the SMBH
mass function of \citet{marconi04}, AGNs hosting SMBHs with $\Mbh
\approx (0.5$--$5) \times 10^{6} \Msun$ are a factor $\approx 1.5$
more abundant than those with $\Mbh \approx (0.5$--$5) \times 10^{7}
\Msun$. Hence, based on this simplistic formalism, we estimate that
approximately half of all Compton-thick AGNs with $L_{\rm X} \goa
10^{42} \ergps$ may contain SMBHs with $\Mbh \approx (0.5$--$5) \times
10^{6} \Msun$ which are not included in our parent sample.

Obscuration and host-galaxy contamination may further prevent us from
identifying all Compton-thick AGNs in our considered volume. Obscured
AGNs can be mis-classified when the host-galaxy over-shines the
nuclear emission. In the absence of extinction, a NL-AGN with $L_{\rm
  X} \approx 2 \times 10^{42} \ergps$ can be almost totally diluted by
a star-formation rate of $10 \Msun$~yr$^{-1}$ (i.e., $\goa 95$~percent
of the observed $H \beta$ emission is produced in H{\sc ii} regions;
e.g., \citealt{yan10}). The contribution of these sources to our
observed space density is difficult to quantify. However, it is
predicted that as many $\approx 50$~percent of AGNs may show no
evidence for AGN activity in their optical spectroscopy (e.g.,
\citealt{maiolino03}; Goulding \& Alexander 2009), and would therefore
not be included in our optically selected AGN sample. Allowing for the
incompleteness within our optical parent sample and the possibility
that many more of the AGNs studied here may be Compton thick, we
suggest that our derived space density can be broadly consistent with
the XRB models.

\subsection{The mean growth rate of Compton-thick AGNs at $z \sim 0.1$}
\label{subsec:growth_cthick_agns}

Some theoretical models predict that Compton-thick AGNs may harbour
SMBHs which are undergoing an evolutionary phase of rapid growth
(e.g., \citealt{Fabian99,Granato06,Hopkins08}). In this section we
consider the implied Eddington ratios ($\eta \sim L_{\rm AGN} / L_{\rm
  Edd}$; where $ L_{\rm Edd} \approx 1.26 \times 10^{38} (\Mbh /
\Msun) \ergps$) for the Compton-thick AGNs identified in our sample
with publicly available black-hole mass ($\Mbh$) estimates. Stellar
velocity dispersion measurements have been computed for 13 of the 14
AGNs in our sample, at least five of which we conservatively identify
as Compton-thick AGNs. These measurements are publicly available in
the MPA-JHU release of SDSS-DR7 and are derived from the fitting of
stellar population synthesis models to the SDSS 1-D
spectra.\footnote{The MPA-JHU SDSS catalogue is maintained by a large
  collaboration of SDSS researchers and is a complementary dataset
  providing additional information for the SDSS-DR7 data-release
  including measurements of velocity dispersions, stellar masses,
  star-formation rates etc. It is available at
  http://www.mpa\-garching.mpg.de/SDSS/} Using the $M$--$\sigma$
relation of \citet{gebhardt00} we convert the stellar velocity
dispersions to $\Mbh$ (see Column 7 of Table~\ref{tab:srce_props}) in
order to calculate $L_{\rm Edd}$ for these sources. The median SMBH
mass for our sample is $\Mbh \approx 3 \times 10^7 \Msun$ (i.e., these
AGNs host SMBHs which are similar to those identified in the optical
study of Heckman et~al. 2004).

In order to estimate $\eta$ for our Compton-thick AGNs, we use $L_{6
  \mu m}$ as a proxy for $L_{\rm AGN}$ and we assume the bolometric
corrections of \citet{marconi04}. The use of the $6 \um$ continuum
emission to infer $L_{\rm AGN}$ has the advantage that it is an
independent measure of the intrinsic luminosity of the AGN, whilst the
NL \oiv emission arises from a similar region to that of [O{\sc iii}]
which was used for the selection of the sources considered here.

Twelve of our 14 sources have both $\Mbh$ estimates and $6 \um$
measurements. We find that our sample of AGNs are spread over a
wide-range of Eddington ratio, $\eta \approx 0.002$--0.3
(median~$\approx 0.014$; see Column 11 of Table 2); the five
Compton-thick AGNs are at systematically higher Eddington ratios,
$\eta \approx 0.01$--$0.3$ (median~$\approx 0.2$).\footnote{We note
  that when using \oiv emission to infer $L_{\rm AGN}$, to
  first-order, we find a similar result (i.e., systematically higher
  Eddington ratios for the Compton-thick AGNs).} Similarly, we find
that the well-studied local Compton-thick AGNs (Circinus, Mrk~3,
NGC~1068 and NGC~6240) have a similar range in Eddington ratio, $\eta
\approx 0.002$--1 (median~$\approx 0.2$). It is important to note the
large uncertainties involved with calculating bolometric luminosities
and subsequent Eddington ratios; the large scatter in the X-ray--$6
\um$ relation combined with a possible Eddington ratio dependent
bolometric correction (e.g., \citealt{vasudevan07}) could yield an
uncertainty factor of the order $\goa 10$ for the highest Eddington
ratio sources. None of the AGNs in our sample appear to be
Eddington-limited on the basis of the $6 \um$ luminosity, and any
uncertainties would apply equally to all of the AGNs considered here,
hence our finding of systematically higher Eddington ratios for
Compton-thick AGNs, to first-order, appears to be relatively
robust. However, we suggest that this result may be driven by our
selection of [O{\sc iii}]-bright AGNs as well as our sensitivity
towards the identification of Compton-thick AGNs. For example, with
deeper X-ray data we may identify further Compton-thick AGNs in our
sample which have lower values of $\eta$, and hence reducing the
median Eddington ratio for the Compton-thick AGN subsample.

By comparison, for the total population of Type-2 AGNs identified from
optical emission-line diagnostics in the SDSS, \citet{heckman04} find
that based on the use of [O{\sc iii}] emission to infer $L_{\rm AGN}$,
$< 0.5$~percent of Type-2 AGNs hosting SMBHs with $\Mbh \approx 3
\times 10^7 \Msun$ are accreting above $\eta \approx 0.1$; by
contrast, we find that $\goa 30$~percent of our sample have $\eta \goa
0.1$. A particular advantage to a direct comparison with the study of
Heckman et~al. (2004) is that selection processes and biases are
likely to be identical between both studies. Hence, these results
suggest that, on average, the Compton-thick AGNs identified here may
harbour some of the most rapidly growing black holes in the nearby
Universe. This would further suggest that not taking account of
Compton-thick AGNs in deep-field X-ray surveys may exclude the most
rapid growth phases of SMBHs, as predicted by many theoretical models
(e.g., \citealt{Fabian99,Granato06,Hopkins08}).


\section{Conclusions}

We have presented a sample of 14 local ($z \sim 0.03$--0.2) X-ray
undetected optical AGNs selected from the large overlap region between
the SDSS-DR7 and 2XMMi catalogues. These sources were selected as
candidate Compton-thick AGNs on the basis of their X-ray--\oiii
emission line ratios (i.e., $f_X / f_{\rm [OIII]} < 1.0$; e.g.,
Bassani et~al. 1999; see Section~\ref{sec:g10b_sampleselect}). We have
employed a suite of optical (e.g., \oiii emission-line) and mid-IR
(e.g., \oiv emission-line; $6 \um$ AGN continuum) diagnostics to infer
the intrinsic AGN luminosity in these sources. Assuming any deficit in
X-ray flux compared to these estimates is due to Compton-thick
absorption, we assess the ubiquity of Compton-thick AGN activity in
the nearby Universe. Our main findings are the following:

\begin{enumerate}
\renewcommand{\theenumi}{(\arabic{enumi})}

\item {Using {\it Spitzer}-IRS low resolution spectroscopy, we find
  that six of our 14 candidate Compton-thick AGNs have $\goa 3 \sigma$
  detections of \nev and all 14 have $\goa 3 \sigma$ detections of
  [O{\sc iv}]. We performed mid-IR spectral decompositions of our
  sample to establish $6 \um$ AGN continuum luminosities. Using
  established X-ray to mid-IR continuum and emission-line
  relationships, we infer the intrinsic X-ray luminosity of these AGNs
  and conservatively find that 6/14 ($\approx 43 \pm 21$~percent) of
  the sources in our sample appear to be heavily obscured with $N_H
  \goa 1.5 \times 10^{24} \pcmsq$ (i.e., are Compton-thick AGNs). See
  sections~\ref{subsec:spitz_emiss_line_fluxes},~\ref{subsec:spitz_spec_decomps}
  and~\ref{subsec:intrinsic_nh_from_oiv}.}
\item {We used our results to infer the ubiquity of Compton-thick AGNs
  in our SDSS--2XMMi parent sample. We predict that on the basis of
  the analyses presented here that at least $\goa 20$~percent of the
  334 optical Type-2 AGNs in the SDSS-DR7 at $z \sim 0.03$--0.2 are
  obscured by Compton-thick material. This implies a space-density of
  log$(\Phi) \goa -4.9$~Mpc$^{-3}$ for Compton-thick AGNs with $L_X
  \goa 10^{42} \ergps$ at $z \sim 0.1$, which we suggest may be
  consistent with the number density of Compton-thick AGNs predicted
  by XRB synthesis models when accounting for all of our sample
  selection biases. See sections~\ref{subsec_intrinsic_nh_from_6um}
  and \ref{subsec_space_dens_cthick_agn}.}
\item{We establish that the Compton-thick AGNs identified in our
    sample appear to be rapidly accreting. Using $6 \um$ continuum
    luminosity to infer the $L_{\rm AGN}$ and the stellar velocity
    dispersion to estimate $\Mbh$, we find systematically higher
    Eddington ratios for the most conservatively identified
    Compton-thick AGNs ($\eta \approx 0.2$). By comparison to studies
    of local Type-2 AGNs with similar SMBH masses ($\Mbh \approx 3
    \times 10^7 \Msun$ e.g., Heckman et~al. 2004), we find that
    Compton-thick AGNs selected in the SDSS may harbour some of the
    most rapidly growing black holes in the nearby Universe ($z \sim
    0.1$). However, this result may be driven by our Compton-thick AGN
    identification process. See
  section~\ref{subsec:growth_cthick_agns}.}

\end{enumerate}

In summary, we have established that approximately half of these
sources have optical and mid-IR AGN indicators consistent with their
observed X-ray emission being heavily obscured by Compton-thick
material. Indirect multi-wavelength analyses, such as those employed
here, are currently the most practical and reliable technique to
identify Compton-thick AGNs which are 2--3 orders of magnitude further
down the $L_X$--$z$ plane than can be achieved using X-ray
spectroscopy alone. Using the next generation of X-ray satellites
(e.g., {\it NuStar}; {\it IXO}; {\it WFXT}), high-quality X-ray
spectroscopy and $E>10$~keV detections will allow us to directly and
unambiguously identify which of these sources are Compton-thick AGNs.

\section*{Acknowledgments}

We thank the anonymous referee for their considered and thorough
report which has improved the quality of this paper. We would like to
acknowledge useful conversations with J Geach and T Roberts. We thank
R Gilli and E Treister for kindly providing the XRB space density
tracks. We also thank the Science \& Technologies Facilities Council
(ADG; RCH), the Royal Society (DMA) and the Leverhulme Trust (DMA;
JRM) for funding. This research has made use of the Sloan Digital Sky
Survey data archive and the NASA {\it Spitzer} Space Telescope which
is operated by the Jet Propulsion Laboratory, California Institute of
Technology under contract with the National Aeronautics and Space
Administration. This research has also made use of data obtained from
the Leicester Database and Archive Service at the Department of
Physics and Astronomy, Leicester University, UK.

\bibliography{bibtex1}

\bsp 

\label{lastpage}

\end{document}